# Hydrodynamic magnetotransport in two-dimensional electron systems with macroscopic obstacles

P.S. Alekseev and A. P. Dmitriev

*Ioffe Institute, Politekhnicheskaya 26, Saint Petersburg 194021, Russia*

**Abstract**

In high-quality conductors, the hydrodynamic regime of electron transport has been recently realized. In it, the inter-particle interaction leads to formation of a viscous electron fluid. In this work we theoretically investigate the magnetotransport properties of a viscous electron fluid in samples with electron-impermeable obstacles. We assume that their size is much smaller than the average distance between them and for simplicity consider the disks to be the same with radius $R$. We use the two approaches to describe the fluid flow. The first one is based on the equations of hydrodynamics of a charged fluid, which assume, as we know, that the kinetic equation takes into account the two harmonics of the electron distribution function. The second approach is based on the equations that are obtained by taking into account three harmonics of the distribution function ("quasi-hydrodynamics"). Herewith the condition $l_2 << R$ is assumed to be satisfied, where $l_2$ is the relaxation length of the second harmonic of the distribution function. Within the hydrodynamic approach, we consider the cases of the rough and the smooth edges of the disks, on which the electron scattering is diffusive or specular, respectively. For these two systems, we derive expressions for the components of the resistivity tensor. The longitudinal magnetoresistivity turns out to be strong and negative, the same for both rough and smooth discs edges to within small corrections. For rough discs, the Hall resistivity is equal to its standard value, stemming just from the balance of the Lorenz magnetic force and the electric force in the direction perpendicular to the flow. For smooth discs the Hall resistance acquire a small correction to the standard value, proportional to the Hall viscosity. In the quasi-hydrodynamic approach, we considered the case of smooth discs and small magnetic fields. When the inequality $\sqrt{l_2 l_3} << R$ is fulfilled, hydrodynamics approach is applicable, whereas under the conditions $\sqrt{l_2 l_3} >> R$ the flow is forms which is substantially different from the hydrodynamic one (here $l_3$ are the relaxation length of the third harmonic of the distribution function). In the last regime, the longitudinal resistivity does not depend on the relaxation length $l_2$ and the correction to the standard Hall resistivity does not depend on both lengths $l_2$ and $l_3$, but depends only on the concentration and size of the disks. We compare the results of the hydrodynamic calculation of the longitudinal resistance with the experimental data on magnetotransport in high-quality GaAs quantum wells with macroscopic defects. A good agreement of theory and experiment evidences in favor of the realization of the hydrodynamic transport regime in suchsystems.

# I. Introduction

### 1. Hydrodynamics of viscous electron fluid in solids

Frequent electron-electron collisions in high-quality conductors can lead to formation of a viscous fluid and realization of the hydrodynamic regime of charge transport [1]. In such systems, flows of the electron fluid are space-inhomogeneous and determined by geometry of a sample. This transport regime was recently reported for high-quality graphene [2-6], layered metal PdCoO$_2$ [7], Weyl semimetal WP$_2$ [8], and GaAs quantum wells [9-24]. Formation of the electron fluid was detected by a specific dependence of the resistance on the sample width [7,24], by observation of the negative nonlocal resistance [2,3,15,22], by the giant negative magnetoresistance [8-14,16,17,23,24], and by the magnetic resonance at the double cyclotron frequency [18-21,24].

There are various types of samples, differentin their geometry,where the viscous flows of the electron fluid were reported. The simplest one is the flat geometry of the Poiseuille flow in a long narrow samples. Such samples were studied in Refs. [2-8,24]. In this case the flow is quasi-one-dimensional: its profile depends only on the coordinate perpendicular to the longitudinal edges of the sample. Hydrodynamic flow of another type is formed in a high-quality sample with localized macroscopic defects, on average homogeneously distributed and impermeable to electrons. Besides this, the hydrodynamic electric transport has been studied in the samples of a variety of complex geometries and with complex arrangements of the electric contacts. For example, in Refs. [25,26] experimental and theoretical studies of the flow of the electron fluid in a long sample with one obstacle inside the bulk of the sample were performed.

Samples with macroscopic defects were studied in Ref. [13,27]. In [13] GaAs quantum wells samples with "oval defects", which appeared in the growth process, were examined. The electron mean free path related with scattering on these defects was determined from the sample resistance in zero magnetic field. In classically strong magnetic fields, the giant negative magnetoresistance, which evidences the formation of the viscous electron fluid [17], was observed in Ref. [13]. In Ref. [27] a set of samples of GaAs quantum wells were fabricated in which localized macroscopic obstacles of different densities were made using electron beam lithography and subsequent reactive ion etching. Extensive magnetotransport measurements of those samples were performed, in which various types of the giant negative magnetoresistance were observed.

A first theory of the hydrodynamic charge transport of the two-dimensional electron fluid in a sample with localized defects was constructed in Ref. [28]. Recently, in Ref. [29] a theory describing the crossover between the Ohmic and the hydrodynamic regimes in such systems with increase of the inter-particle scattering rate was constructed. In Refs. [28,29] and other previous works only electron fluid flows in the absence of magnetic fields were studied.

## 2. Flows of ordinary viscous fluids via porous media

In fact, flows of uncharged fluids through an array of obstacles have been systematically studied in ordinary hydrodynamics many years ago. A simplest example of such systems is a flow of water through an array of rocks in a mountain river. A more general example is a flow of a fluid via a porous media formed by randomly placed obstacles. The example of such systems in chemistry is the so called "packed bed", which is used to improve contact between two phases, a solid and liquid, in a chemical process.

Simplest qualitative description of such system is the Kozeny-Carman equation [30,31]. It models the fluid flow in a sample of porous media as laminar fluid flow in a collection of curving passages crossing the packed bed. For each passage, the Poiseuille law is used to describe the laminar fluid flow in each section of the passage. Then the averaging of these results is performed to calculate the whole flow and the pressure drop in a sample.

There are two more rigorous analytical approaches for the calculation of relation between the pressure drop and the flow in systems where the average distance between obstacles far exceeds their size.

The first one, known as the Brinkman approach, is the effective media approximation. In it, only the flow near one (any) obstacle is explicitly considered, while the influence of other obstacles is taken into account by introducing the term $-\mathbf{V}(\mathbf{r})/\tau$ into the Navier-Stokes equation [32]. Herewith, at a large distance from the considered obstacle, the hydrodynamic velocity $\mathbf{V}(\mathbf{r})$ is considered fixed and equal to the average flow velocity in the sample. So the problem of the flow via an array of obstacles is reduced to the problem of a flow around a single obstacle immersed in a dissipative medium, the parameters of which are calculated in a self-consistent way. The microscopic derivation of the Brinkman equation and its corrections is described in Refs. [33,34].

The second approach is the so called cell model. In it, the hydrodynamic Stokes problem is solved for one obstacle with boundary conditions on the obstacle boundary and on an imaginary cell boundary. Boundary conditions on the obstacle edges can be of two types: the Kuwabara conditions for sticky disks [35] and the Happel condition for mirror disks [36,37]. The Happel condition requires the tangential component of the stress tensor to vanish at the disk boundary, while the Kuwabara condition requires zeroing the full hydrodynamic velocity. As for the cell boundary, it is assumed that the hydrodynamic velocity on it is equal to the average velocity in the sample.

There is also a combined approach [38]: a boundary condition on an imaginary cell boundary matches the solution of the Stokes problem inside the cell (solution of usual Navier-Stokes equation) with the solution of the Brinkman problem with the Drude-like friction term $-\mathbf{V}(\mathbf{r})/\tau$ around the cell. The continuity of the velocity, pressure and stress tensor at the cell boundary is required.

Further development of theory leads to many quantitative, detailed results. In particular, it became possible to obtain relationships between the parameters of the effectivemedia and of the flow in order to explain specific experimental data (see, for example, works [39,40]).

In Ref. [17] has recently been proposed a simple qualitative description of the flow of a viscous 2D electron fluid in magnetic field in a sample with macroscopic obstacles, having rough boundaries. An estimate for the sample resistance was derived for the case when the size of the obstacles is of the same order of magnitude as the distances between them. Consideration was performed by a method analogous to the simplest Kozeny-Carman method of description of flows an uncharged fluid via a porous media. The sample resistance turns out to be proportional to the diagonalviscosity, that leads to a strong negative magnetoresistance, similar to the one in a Poiseuille flow.

**3. Hall effect in electron hydrodynamics**

In studies of the hydrodynamic regime of electron transport, the Hall effect was of great interest. It was believed that the Hall voltage in such systems consists of two contributions: the main ``standard'' contribution, associated with the balance of the Lorentz force and electric force, and a contribution arising due to the term of the off-diagonal ``Hall'' viscosity in the Navier-Stokes equation. In Refs. [4,16] it was reported about the measurements of the Hall resistance in samples in which two-dimensional electrons form a viscous fluid. These experiments were performed on different materials (grapheme and GaAs quantum wells) and for samples of different geometries, but their results turned out to be rather similar: the Hall resistance has an additional size-dependent contribution to its standard contribution. The value of the size-dependent contribution in Ref. [4] was directly related to the coefficient of the Hall viscosity.

In Ref. [41] magnetotransport of interacting two-dimensional electrons in long samples with rough edges was theoretically studied. Using the numerical solution of the kinetic equation, the longitudinal and the Hall resistances were calculated for parameters corresponding to both the ballistic and the hydrodynamic regimes of transport. In particular, it was shown that for the samples in which the hydrodynamic regime is realized (the mean free path relative to inter-particle collisions is much less than the sample width), the Hall resistance deviates from the standard value by a small negative value.

In Refs. [42-45] the Hall effect was theoretically studied for systems of interacting electrons in long samples in the ballistic regime and in the transition subregimes between the ballistic and the hydrodynamic regimes. A variety of anomalies were predicted: for example, a giant value of the Hall resistance in the ballistic point [44] and a kink and other singularities in the longitudinal and the Hall resistance in the transition region [45]. Some of these features have been previously discovered in numerical theory [41].

In Ref. [46] the influence of the near-edge layers of the Poiseuille flow on the Hall effect was studied for a long sample with rough edges. In the layers of the widths of the order of the interpaticle scattering length, a half of electrons is reflected from an edge of the sample. Therefore, the flow in such layers is semi-ballistic. It is described by the kinetic equation, in contrast to the rest bulk part of the sample, in which the hydrodynamic equations are applicable. The description of the near-edge layers was carried out in [46] according to a method similar to the one developed in Refs. [47,48] for a Poiseuille flow. It was shown in Ref.[46] that the contribution of the near-edge semi-ballistic layers to the Hall resistance is significant: it is of the same order of magnitude as the contribution from the bulk Hall viscosity term.

### 4. Subject of this work

In this work we develop a theory of the flow of a two-dimensional electron fluid in high-quality samples with rare macroscopic obstacles ("disks") in a perpendicular magnetic field. Throughout the paper it is assumed that the free path of electrons with respect to electron-electron collisions (more precisely, the relaxation length of the shear stress in the fluid) is much smaller than obstacle size. We develop two approaches.

The first one is hydrodynamic. We consider a long sample with discs and solve the Navier-Stokes equation for two-dimensional fluid hydrodynamic, accounting the diagonal and the non-diagonal (Hall) viscosities. Specific calculations were performed within the effective medium method, which we generalized to the case of a charged fluid in a magnetic field. Following to Refs. [32-34], we calculate average characteristics of the flow in the whole sample by consideration of the flow around some disk immersed in an effective medium. The last one consists of the fluid and all other discs and provides effective resistance for the flow far from a chosen disc.

We find the longitudinal and Hall resistivity's within such model. Owing to the magnetic field dependence of the diagonal viscosity $\nu$, a strong negative magnetoresistance, similar to one in the Poiseuille flow [17], arises. With small corrections, the longitudinal resistance does not depend on the boundary conditions at the edges of the disks. The Hall resistance depends on the type of disc edges. In the case of rough edges it is exactly equal to the standard Hall resistance corresponding to the balance between the magnetic Lorenz force and the electric force in the bulk of the fluid. On the contrary, in the case of the smooth disk edges, a correction to the standard Hall resistance which is proportional the Hall viscosity arises.

The second approach goes beyond hydrodynamics. The equations of hydrodynamics are derived from the kinetic equation assuming that all harmonics of the distribution function starting from the third one relax instantaneously. In fact, the relaxation time of the third harmonic is of the same order as the relaxation time of the second harmonic or even far exceeds it [49]. So the question arises, how would the predictions of the purely hydrodynamic theory change, if the three harmonics were included in the calculations? Besides, from the analogy with the consideration of Ref. [46], it is obvious that a thin semi-ballistic layer around the obstacles edges is formed. In this layer, the hydrodynamic approximation is not applicable and higher harmonics of the distribution function should be taken into account. Approximation of three harmonics is a step in this direction. We derive a system of two-dimensional "quasi-hydrodynamic" equations of the electron fluid from the kinetic Boltzmann equation in the three-harmonic approximation and solve it using the effective medium method.

It turned out that in addition to the spatial scales $l_2$, $R$ and $\lambda_\tau = \sqrt{\nu\tau}$, being characteristic for the hydrodynamic approximation and satisfying the inequality $l_2 \ll R \ll \lambda_\tau$, another scale, $\lambda \sim \sqrt{l_2 l_3}$, arises. We demonstrated that the last scale, depending on the relaxation length $l_3$, can be greater than $l_2$ and $R$, but is always smaller than $\lambda_\tau$. If the inequalities $\lambda \ll R$, $l_3 \ll R$ are fulfilled, then thin semi-ballistic layers around the discs have the width $\lambda$ and the resistivity tensor is equal to the hydrodynamic result with an accuracy of small corrections. If $\lambda \ll R \ll l_3$, then the longitudinal resistivity still coincides with its hydrodynamic value, while the correction to the standard Hall resistivity turns out to be non-hydrodynamic. Finally, if $R \ll \lambda \ll l_3$, then the flow regime becomes, as mentioned above, non-hydrodynamic one: the longitudinal resistivity do depend not on the shear stress relaxation length $l_2$, by on the length $l_3$ controlling the relaxation of the ballistic contribution. Herewith the Hall resistance does not depend on both lengths $l_2$ and $l_3$ and is determined only by the properties of obstacles.

Below, for convenience, we describe the structure of the rest part of the article.

In Section II.1, starting from Navier-Stokes equations in a linear approximation, we present general rigorous statements and derive equations, which are the basis of our further calculations.

In Section II.2, we briefly describe the effective medium method as applied to the hydrodynamic equations of the two-dimensional electron fluid.

In Section II.3 we perform exact calculations of the velocity field and the resistivity tensor for the flows via arrays of rough (diffusely scattering electrons) and smooth (reflecting them mirror-like) disks. In both cases the longitudinal resistivity was found to be the same up to small corrections. Its main part is and proportional to the longitudinal viscosity and inversely proportional to the logarithm of the ratio of the average distance between the disks and the disk radius. Due to the dependence of the viscosity coefficient on magnetic field, this leads to the giant negative magnetoresistance similar to the one in the Poiseuille problem [17]. In the case of rough discs, the Hall resistance coincides with its standard value, whereas in the case of smooth discs there is a small correction to the standard value, being proportional to the Hall viscosity.

In Section II.4 we calculate fluctuations of the components of the resistivity tensor due to irregularities in the arrangement of the disks. We show that the resistivity fluctuations are much smaller than their mean values.

In Section II.5, we present and briefly discuss the velocity profile in the vicinity of a disk.

In Section III, we derive quasi-hydrodynamic equations based on the three-harmonic approximation for the distribution function and solve them for the case of smooth discs within the effective medium method.

In Section IV we compare our results for the longitudinal resistance with experimental data on magnetoresistance of Ref. [27] for high-quality GaAs quantum well samples in which localized macroscopic obstacles of different densities were made by ion eatching. The observed giant negative magnetoresistance of different amplitudes is compared with our results for the cases of the flow via array of discs with rough or smooth edges. A good agreement between theory and experiment is demonstrated, however. To describe the experimental data we used only one fitting parameter, the shear stress relaxation rate $1/\tau_2$. It allow to describe both the shape, the width, and the amplitude of the observed magnetoresistance curves.

In Conclusion we sketch all the obtained results and discuss their importance and perspectives.

## II. Hydrodynamic approach

### 1. Basis equations

We consider a two-dimensional long sample of length $L$ and width $W \ll L$, located in the plane $(x, y)$ and oriented along the axis $x$ (see Fig. 1). A homogeneous magnetic field $B$ is applied in the direction of the axis $z$. It is assumed that the sample contains a degenerate electron gas and that interelectron collisions are the dominant mechanism of electron scattering.

We also suppose that the sample contains macroscopic electron-impermeable obstacles, which we will assume for simplicity to be discs with the radius $R$. The concentration of the discs $n_D$ is assumed to be small enough, so that the inequality $n_D R^2 \ll 1$ is fulfilled. Throughout this article, the inequality $l_2 \ll R$ will be assumed to be fulfilled, which will allow us to use the equations of hydrodynamics to describe the motion of electron gas. All specific results will be obtained in the thermodynamic limit, $S = LW \to \infty$ provided $n_D = \text{const}$.

Our problem is to find the distribution of flow of the electron fluid in the sample and the resistivity tensor of the sample $\rho_{\alpha\beta}$.

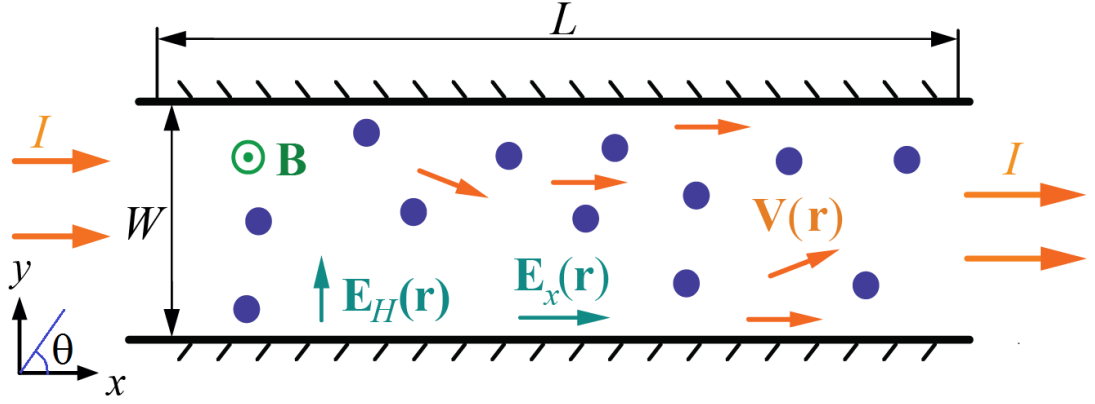

**Fig. 1. Long sample with discs and 2D electron Fermi gas.**

We also assume that the perturbation of the electron system caused by the time-independent voltage applied to the sample is small, that allows us to use the linearized Navier-Stokes and continuity equations to solve the problem:

$$\text{div } \mathbf{V} = 0, \quad e\mathbf{E} + m\omega_c[\mathbf{V} \times \mathbf{e}_z] - m\nu\Delta\mathbf{V} + m\nu_H[\Delta\mathbf{V} \times \mathbf{e}_z] = 0 \cdot \qquad (1)$$

Here $\mathbf{V}(\mathbf{r})$ is the velocity of the electron fluid at the point $\mathbf{r}$, the symbol $\mathbf{E}(\mathbf{r})$ means the gradient of the electrochemical potential $\Psi(\mathbf{r})$ taken with a minus sign, $\mathbf{e}_z$ is a unit vector along the z axis, $\nu = \nu_0/(1+\beta_2^2)$ is the diagonal viscosity coefficient depending on magnetic field, $\beta_2 = 2\omega_c\tau_2$, $\nu_H = \beta_2\nu$ is the Hall viscosity coefficient, $\omega_c$ is the cyclotron frequency, and $\tau_2$ is the relaxation time of the second harmonic of the distribution function, or, what is the same, the relaxation time of the shear stress. The distribution of the discs is considered to be random and homogeneous, on the scales larger than the average distance between the discs, so that in sufficiently large samples the electric field and the electron flow are homogeneous on the same scales.

The first of equations (1) allows us to express the hydrodynamic velocity and the vorticity through the flow function $\psi$:

$$V_x = -\frac{\partial \psi}{\partial y}, \quad V_y = \frac{\partial \psi}{\partial x}, \quad \Omega = \frac{\partial V_y}{\partial x} - \frac{\partial V_x}{\partial y} = \Delta\psi \cdot \qquad (2)$$

Introducing the function $e\tilde{\Psi} = e\Psi - m\nu_H\Omega - m\omega_c\psi$, let us rewrite the Navier-Stokes equations in the form:

$$\frac{\partial \tilde{\Psi}}{\partial x} = -\frac{m\nu}{e}\Delta V_x, \quad \frac{\partial \tilde{\Psi}}{\partial y} = -\frac{m\nu}{e}\Delta V_y, \tag{3}$$

or, equivalently:

$$\frac{\partial \tilde{\Psi}}{\partial x} = \frac{m\nu}{e}\frac{\partial \Omega}{\partial y}, \quad \frac{\partial \tilde{\Psi}}{\partial y} = -\frac{m\nu}{e}\frac{\partial \Omega}{\partial x}. \tag{4}$$

By differentiating the first of these equations by $y$, and the second by $x$, and subtracting the results from each other, we obtain a simple equation for the vorticity $\Delta\Omega = 0$. By virtue of Eq. (2), it is equivalent to the biharmonic equation:

$$\Delta^2 \psi = 0. \tag{5}$$

By integrating equations (4) over the flow area and dividing the result by the sample area $S$, we obtain:

$$\frac{1}{S}\int_0^W [\tilde{\Psi}(L,y) - \tilde{\Psi}(0,y)]dy = \frac{R}{S}\sum_k \oint_{\Gamma_k} \tilde{\Psi}\cos\theta_k d\theta_k - \frac{m\nu R}{eS}\sum_k \oint_{\Gamma_k}\Omega\sin\theta_k d\theta_k + \frac{m\nu}{eS}\int_0^L [\Omega(x,W) - \Omega(x,0)]dx, \tag{6}$$

$$\frac{1}{S}\int_0^L [\tilde{\Psi}(x,W) - \tilde{\Psi}(x,0)]dx = \frac{R}{S}\sum_k \oint_{\Gamma_k} \tilde{\Psi}\sin\theta_k d\theta_k + \frac{m\nu R}{S}\sum_k \oint_{\Gamma_k}\Omega\cos\theta_k d\theta_k - \frac{m\nu}{eS}\int_0^W [\Omega(L,y) - \Omega(0,y)]dy$$

The symbol $\oint_{\Gamma_k} ..d\theta_k$ means integration over the edge of the disk with the number $k$. The lines $x=0$ and $x=L$ corresponds to the source and drain, respectively. The lines $y=0$ and $y=W$ to the lower and upper edges of the sample in the direction of the ordinate axis. The angle $\theta_k$ in each summand is counted from the abscissa axis.

From Eq. (5) and the equality $\tilde{E}_\theta = -\partial\tilde{\Psi}/R\partial\theta = m\nu\partial\Omega/e\partial r$, which follows from (4), we get:

$$\frac{1}{S}\int_0^W [\tilde{\Psi}(L,y) - \tilde{\Psi}(0,y)]dy = \frac{m\nu R}{eS}\sum_k \oint_{\Gamma_k}\left(R\frac{\partial\Omega}{\partial r} - \Omega\right)\sin\theta d\theta + \frac{m\nu}{eS}\int_0^L [\Omega(x,W) - \Omega(x,0)]dx, \tag{7}$$

$$\frac{1}{S}\int_0^L [\tilde{\Psi}(x,W) - \tilde{\Psi}(x,0)]dx = \frac{m\nu R}{S}\sum_k \oint_{\Gamma_k}\left(-R\frac{\partial\Omega}{\partial r} + \Omega\right)\cos\theta d\theta - \frac{m\nu}{eS}\int_0^W [\Omega(L,y) - \Omega(0,y)]dy$$

Transferring now the functions $-(m\omega_c/e)\psi$ and $-(m\nu_H/e)\Omega$, included in $\tilde{\Psi}$, from the left parts of these equations to the right ones and introducing the notations $U_x$ and $U_y$ for the differences of electrochemical potentials (voltages), we obtain

$$\frac{U_x}{L} = \frac{m\nu R}{eS}\sum_k \oint_{\Gamma_k}\left(R\frac{\partial\Omega}{\partial r} - \Omega\right)\sin\theta d\theta + \frac{m\omega_c}{e}\overline{V}_y + \frac{m\nu}{eS}\int_0^L [\Omega(x,W) - \Omega(x,0)]dx + \frac{m\nu_H}{eS}\int_0^W [\Omega(L,y) - \Omega(0,y)]dy, \tag{8}$$

$$\frac{U_y}{W} = -\frac{m\nu R}{eS}\sum_k \oint_{\Gamma_k}\left(R\frac{\partial\Omega}{\partial r} - \Omega\right)\cos\theta d\theta - \frac{m\omega_c}{e}\overline{V}_x - \frac{m\nu}{eS}\int_0^W [\Omega(L,y) - \Omega(0,y)]dy + \frac{m\nu_H}{eS}\int_0^L [\Omega(x,W) - \Omega(x,0)]dx$$

In Appendix A we present details of derivation of Eqs. (8), in particular, we show, how in these equations appear the velocities averaged over the flow region $\overline{V}_x$ and $\overline{V}_y$.

The integrals along the edges of the sample in the thermodynamic limit turn to zero. Indeed, the increase of the vorticity $\Omega$ with the increase of the sample size would mean an increase in velocity with increasing size, which is obviously not the case. Thus we obtain:

$$-m\omega_c \overline{V}_y + \frac{eU_x}{L} - \frac{m\nu R}{S} \sum_k \oint_{\Gamma_k} \left( R\frac{\partial \Omega}{\partial r} - \Omega \right) \sin\theta \, d\theta = 0,$$

$$m\omega_c \overline{V}_x + \frac{eU_y}{W} + \frac{m\nu R}{S} \sum_k \oint_{\Gamma_k} \left( R\frac{\partial \Omega}{\partial r} - \Omega \right) \cos\theta \, d\theta = 0$$
(9)

In other words, the transition to these equations is possible only in the case of sufficiently large samples when the contribution to the resistance associated with the edges is small compared to the contribution from the inner region. Equations (9) are the basis for further calculations. Another form of writing these equations, using the momentum flux density tensor, is presented in Section 3, Eqs. (46) - (48).

It is important that equations (5) and (9) do not contain terms with Hall viscosity. From this fact the following theorem. In the case when the Hall viscosity $\nu_H$ is absent in the boundary conditions on the disc edges, the vorticity $\Omega$ also does not depend on the value $\nu_H$ according to Eqs. (2) and (5). Therefore the components of the resistivity tensor also do not depend on $\nu_H$. In Appendix B we show that, in fact, this Statement is true for obstacles of any shape. It is not difficult to show that the voltages on the disks and on the flow region separately contain contributions proportional to $\nu_H$, but in total they exactly compensate each other. It is important for this compensation that the flow bypasses around the disks from all sides. To calculate the integral over the interior of the disk, we should account for the potentiality of the electric field. Namely, the integral of the electric field over the curve that lies inside the disk and connects two points on its edge is equal to the integral over the curve that lies outside the disk and connects the same two points. The statement formulated can be proven just as strongly, but more physically. Namely, it is not difficult to show that the work of the term with the Hall viscosity is zero - the corresponding calculations are given in Appendix C.

From equations (2) and (5) it also follows that the velocity profile in the sample is independent of the magnetic field if the boundary conditions do not depend on it.

We emphasize that both of these statements are true only in the case of a small perturbation, when it is possible to discard all nonlinear contributions.

Note that in a finite-size sample there is a small correction to the resistance containing the Hall viscosity. It is given by the integrals at the edges of the sample. In the absence of discs, this correction is responsible for the deviation of the Hall resistance from its standard value in the Poiseuille flow problem.

We conclude this section by noting that in contrast to the three-dimensional systems with two-dimensional symmetry ("columns" along the axis $z$ instead of disks), where in the absence of magnetic field the charges are concentrated at the boundaries of the flow region, in the two-dimensional case the charges creating the electric field in the sample are distributed over its entire surface. This means that a two-dimensional electron fluid, unlike a three-dimensional fluid, cannot be considered fully incompressible. Some perturbations of the 2D electron density are implied in the above consideration as they are responsible for the appearance of the internal electric field, in particular, the Hall field. However, in linear approximation, we can find from equation (2) the internal electric field corresponding to the solution of Eq. (5) for the flow function ψ and the velocity **V(r)** and then, by the electrostatic formulas, we can calculate the charge density, being proportional to the perturbation of the electron concentration.

## 2. Effective medium method

The exact profile of the fluid flow cannot be calculated for a given arbitrary arrangement of disks in a sample. Therefore we propose a method of derivation of mean hydrodynamic equations averaged over the positions of disks, i.e., over the realizations of disorder. In this derivation we will follow works [32, 39, 40], in which the problem of a viscous neutral fluid flowing through a array of randomly placed solid obstacle was considered. It was shown there that, under the condition $n_D R^2 \ll 1$, the problem reduces to describing the fluid flow in a system with one disk immersed in a medium with an effective relaxation time $\tau$ simulating the influence of all other disks. . The results of these works cannot be transferred directly to the case of a charged fluid, but it is possible to use some ideas expressed there.

First of all, we average equations (9) over the positions of disks, assuming their distribution over the sample to be homogeneous on average and neglecting the contribution of disks close to the edges of the sample. The result is:

$$-m\omega_c \overline{V}_y + \frac{eU_x}{L} = m\nu R n_D \oint_\Gamma \left( R \frac{\partial \langle \Omega \rangle_1}{\partial r} - \langle \Omega \rangle_1 \right) \sin\theta d\theta, \quad (10)$$

$$m\omega_c \overline{V}_x + \frac{eU_y}{W} = -m\nu R n_D \oint_\Gamma \left( R \frac{\partial \langle \Omega \rangle_1}{\partial r} - \langle \Omega \rangle_1 \right) \cos\theta d\theta$$

where angle brackets with index 1 mean the average value under the condition that one disk (any disk) is fixed and averaging is performed on the positions of other disks. The values $\overline{V}_x$ and $\overline{V}_y$ will hereafter be considered as given, determined by the given flows of the electron fluid. The voltages are averaged, but we will not put them in angle brackets (without index). The method for finding $\langle \Omega \rangle_1$ is set forth below.

The calculations presented in Appendix D lead us to Eq.

$$-e\mathbf{E} + m\nu \Delta \mathbf{V} - m\nu_H [\Delta \mathbf{V} \times \mathbf{e}_z] - m\omega_c [\mathbf{V} \times \mathbf{e}_z] - m\delta\omega_c [\mathbf{V} \times \mathbf{e}_z] - \frac{m\mathbf{V}}{\tau} = 0. \quad (11)$$

where $\tau$ and $\delta\omega_c$ at $\overline{V}_y = 0$, which is the case in our problem, are given by the expressions

$$\frac{1}{\tau} = \frac{\nu R n_D}{\overline{V}_x} \oint_\Gamma \left( R \frac{\partial \langle \Omega \rangle_1}{\partial r} - \langle \Omega \rangle_1 \right) \sin\theta d\theta, \quad (12)$$

$$\delta\omega_c = \frac{\nu R n_D}{\overline{V}_x} \oint_\Gamma \left( R \frac{\partial \langle \Omega \rangle_1}{\partial r} - \langle \Omega \rangle_1 \right) \cos\theta d\theta$$

Such equation implies that the resistance force of the effective medium can be written in the form:

$$\mathbf{f} = -\frac{m\mathbf{V}}{\tau} - m\delta\omega_c [\mathbf{V} \times \mathbf{e}_z].$$

Let us emphasize that the average local field $\mathbf{E}(\mathbf{r})$ in Eq. (11) includes both the contribution of configurations in which the point $\mathbf{r}$ lies in the flow region and the contribution of configurations in which it appears inside one of the disks for the positions of which the averaging is performed.

Applying the rotor operation to equation (11), we obtain the equation for $\langle \Omega \rangle_1$:

$$\Delta \langle \Omega \rangle_1 - \frac{1}{\lambda_\tau^2} \langle \Omega \rangle_1 = 0, \quad \lambda_\tau = \sqrt{\nu\tau}, \quad \varepsilon^2 = \frac{R^2}{\lambda_\tau^2} \ll 1. \tag{13}$$

The solution of this equation, however, will not help us, since the boundary conditions are not set for vorticity, but for velocity, which is not directly expressed by $\langle \Omega \rangle_1$. Since $\Omega = \Delta \psi$, it follows from Eq. (13) (angle brackets and index 1 we hereafter omit):

$$\Delta^2 \psi - \frac{1}{\lambda_\tau^2} \Delta \psi = 0. \tag{14}$$

After solving this equation, we can find $\mathbf{V}(\mathbf{r})$ and $\Omega(\mathbf{r})$ by Eq. (2).

Because of the invariance of equations (14) with respect to rotations, it is natural to represent their general solution in the one-disk problem as a sum over harmonics $\exp(im\theta)$. The condition for the velocity field to be homogeneous at infinity requires $m = \pm 1$. Therefore we construct the solution being proportional to these $m = \pm 1$ harmonics. The general real solution of equation (14), containing only the first harmonic, is:

$$\psi = 2\operatorname{Re}\{[\alpha\rho + \delta/\rho + \gamma K_1(\varepsilon\rho) + \mu I_1(\varepsilon\rho)]\exp i\theta\}, \tag{15}$$

where α, β, γ, and δ are constants independent on ρ and θ; $\boldsymbol{\rho} = \mathbf{r}/R$; $K_1(x)$ and $I_1(x)$ are modified Bessel functions. The function $I_1(\varepsilon\rho)$ exponentially increases with increasing argument, so we assume $\mu = 0$, the function $K_1(\varepsilon\rho)$, in contrast, exponentially decreases at large distances. Therefore, we have:

$$\psi = 2\operatorname{Re}\{[\alpha\rho + \delta/\rho + \gamma K_1(\varepsilon\rho)]\exp i\theta\}. \tag{16}$$

Since the velocity vector of the fluid at $\rho \to \infty$ tends to a coordinate-independent value $(\overline{V}_x, \overline{V}_y)$, according to Eqs. (2) and (16), we get:

$$\alpha = \frac{R\overline{V}_y}{2} + i\frac{R\overline{V}_x}{2} \equiv \alpha_1 + i\alpha_2. \tag{17}$$

The radial velocity $V_r = -\partial \psi / r \partial \theta$ at the edge of the disk (at $\rho = 1$) is zero. It follows from Eq._(16):

$$\alpha + \delta + \gamma K_1(\varepsilon) = 0, \quad \psi = 2\operatorname{Re}\{[\alpha(\rho - 1/\rho) - \gamma K_1(\varepsilon)/\rho + \gamma K_1(\varepsilon \rho)]\exp i\theta\}. \quad (18)$$

Substituting this expression into the equality $\Omega = \Delta \psi$, we obtain:

$$\Omega(r,\theta) = \frac{\varepsilon^2}{R^2} K_1(\varepsilon \rho)[\gamma \exp(i\theta) + \gamma^* \exp(-i\theta)]. \quad (19)$$

The coefficient $\gamma$ will be found below from the second boundary condition at the edges of the disks separately for the cases of rough and smooth disks. From (12) and (19) it follows

$$\frac{1}{\tau} = \frac{2\pi \nu n_D \varepsilon^3 K_2(\varepsilon)}{R \overline{V}_x} \gamma_2, \quad \delta\omega_c = -\frac{2\pi \nu n_D \varepsilon^3 K_2(\varepsilon)}{R \overline{V}_x} \gamma_1. \quad (20)$$

that allow us to self-consistently find expressions for the parameters of the effective medium $\tau$ and $\delta\omega_c$.

From equations (10) we obtain (recall that $\overline{V}_y = 0$):

$$\frac{U_x}{L} = -\frac{2\pi m \nu n_D \varepsilon^3 K_2(\varepsilon)}{eR} \gamma_2, \quad \frac{U_y}{W} = -\frac{2\pi m \nu n_D \varepsilon^3 K_2(\varepsilon)}{eR} \gamma_1 + m\omega_c \overline{V}_x. \quad (21)$$

From here we find:

$$\rho_{xx} = \frac{2\pi m \nu n_D \varepsilon^3 K_2(\varepsilon)}{e^2 n_0 R \overline{V}_x} \gamma_2,$$

$$\rho_{xy} \equiv \rho_H = \rho_H^0 + \frac{2\pi m \nu n_D \varepsilon^3 K_2(\varepsilon)}{e^2 n_0 R \overline{V}_x} \gamma_1, \quad \rho_H^0 = \frac{B}{en_0 c} \quad (22)$$

In the next section, we will find expressions for the components of the resistivity tensor in the limiting cases of rough and smooth disks.

## 3. Calculation of resistivity tensor

**3.1. Rough disks.** In this case the tangential velocity at the edge of the disk is also zero, so using expression(16) for $\psi$, expression(17) for $\alpha$ (at $\overline{V}_y = 0$) and formula $xK_1'(x) = -K_1(x) - xK_0(x)$, we obtain:

$$\gamma_1 = 0, \quad \gamma_2 = \frac{R \overline{V}_x}{\varepsilon K_0(\varepsilon)}, \quad (23)$$

that yields:

$$\frac{1}{\tau} = \frac{2\pi\nu n_D \varepsilon^2 K_2(\varepsilon)}{K_0(\varepsilon)} \approx \frac{4\pi\nu n_D}{K_0(\varepsilon)}, \quad \delta\omega_c = 0. \tag{24}$$

Here we used the fact that $\varepsilon^2 K_2(\varepsilon) \approx 2$ at $\varepsilon \ll 1$. From equations (24) and (13) it follows the equation on the parameter $\varepsilon$:

$$\varepsilon^2 = \frac{2\pi n_D \varepsilon^2 K_2(\varepsilon) R^2}{K_0(\varepsilon)} \approx \frac{4\pi n_D R^2}{K_0(\varepsilon)}. \tag{25}$$

In the expression for the function $K_0(\varepsilon)$, we should limit ourselves to the main, logarithmic contribution by putting $K_0(\varepsilon) \approx \ln(2/\varepsilon \exp \gamma_E) \approx \ln(1/\varepsilon)$, where $\gamma_E \approx 0.58$ is the Euler constant. The next (power) terms in the expression for the function $K_0(\varepsilon)$ are illegitimate, since the effective medium method we used is valid only in the main order by a small parameter $n_D R^2$. As a result, from Eqs. (24) we find:

$$\frac{1}{\tau} \approx \frac{8\pi\nu n_D}{\ln(A\ln A)} \approx 8\pi\nu n_D \left(\frac{1}{\ln A} - \frac{\ln(\ln A)}{\ln^2 A}\right), \quad A = \frac{1}{8\pi n_D R^2} \gg 1. \tag{26}$$

Finally, we get:

$$\rho_{xx} \approx \frac{8\pi m \nu n_D}{e^2 n_0 \ln(A\ln A)} \approx \frac{8\pi m \nu n_D}{e^2 n_0} \left(\frac{1}{\ln A} - \frac{\ln(\ln A)}{\ln^2 A}\right), \quad \rho_H = \rho_H^0. \tag{27}$$

**3.2. Smooth disks.** In this case, the second boundary condition at the edge of the disk is that the non-diagonal component of the momentum flux tensor is zero:

$$\Pi_{\theta r} = -m\nu\left(\frac{\partial V_\theta}{\partial r} + \frac{1}{r}\frac{\partial V_r}{\partial \theta} - \frac{V_\theta}{r}\right) - 2m\nu_H \frac{\partial V_r}{\partial r} = 0. \tag{28}$$

Note the presence of the Hall viscosity in this boundary condition. Substituting here the expressions for the radial $V_\rho = -\partial\psi/\partial\theta$ and tangential $V_\theta = \partial\psi/\partial\rho$ velocity components, and taking into account the expression (19) for the function $\psi$, we obtain

$$\gamma = \frac{4\alpha(1+i\beta_2)}{\varepsilon[2(1+i\beta_2)K_0(\varepsilon) + \varepsilon K_1(\varepsilon)]}. \tag{29}$$

From this and from formulas (20) it follows

$$\frac{1}{\tau} = \frac{4\pi\nu n_D \varepsilon^2 K_2(\varepsilon)[2K_0(\varepsilon)(1+\beta_2^2) + \varepsilon K_1(\varepsilon)]}{[2K_0(\varepsilon) + \varepsilon K_1(\varepsilon)]^2 + 4\beta^2 K_0^2(\varepsilon)} \approx \frac{4\pi n_D \nu}{K_0(\varepsilon)},$$

$$\delta\omega_c \approx \frac{4\pi n_D \varepsilon^2 K_2(\varepsilon)\varepsilon K_1(\varepsilon)\nu_H}{[2K_0(\varepsilon) + \varepsilon K_1(\varepsilon)]^2 + 4\beta_2^2 K_0^2(\varepsilon)} \approx \frac{2\pi n_D \nu_H}{K_0^2(\varepsilon)(1+\beta_2^2)}. \tag{30}$$

and equation for the parameter $\varepsilon$:

$$\varepsilon^2 = \frac{4\pi n_D R^2 \varepsilon^2 K_2(\varepsilon)[2K_0(\varepsilon)(1+\beta_2^2) + \varepsilon K_1(\varepsilon)]}{[2K_0(\varepsilon) + \varepsilon K_1(\varepsilon)]^2 + 4\beta^2 K_0^2(\varepsilon)} \approx \frac{4\pi R^2 n_D}{K_0(\varepsilon)}, \tag{31}$$

where we have omitted the small correction, since the parameter $\varepsilon$ enters the expressions for the resistivities only under the logarithmic sign. Finally, for the resistivity tensor we have up to the inverse square of the large logarithm:

$$\rho_{xx} \approx \frac{8\pi m \nu n_D}{e^2 n_0}\left(\frac{1}{\ln A} - \frac{\ln(\ln A)}{\ln^2 A} - \frac{\nu^2}{(\nu^2 + \nu_H^2)\ln^2 A}\right), \tag{32}$$

$$\rho_H \approx \rho_H^0 + \frac{8\pi m n_D \nu^2 \nu_H}{e^2 n_0 (\nu^2 + \nu_H^2)\ln^2 A}. \tag{33}$$

Here we see an appearance of the correction to the standard Hall resistance in accordance with Statement 1 in Section II.

Note that the expressions for $\rho_{xx}$ in the cases of rough and the smooth disks differ only by corrections, being small by the inverse square of the large logarithm. If we neglect these corrections, in both cases we will have the following dependence of resistivity on the magnetic field:

$$\rho_{xx} = \frac{m}{e^2 n_0 \tau} \approx \frac{8\pi m n_D \nu}{e^2 n_0 \ln(1/4\pi n_D R^2)},$$

qualitatively coinciding with the experimental one (see Section IV). It is important that the disc radius enters this expression only under the sign of the logarithm. The appearance of the logarithmic dependence here is related to the so-called Stokes paradox.

It can be seen from the expression for $\rho_{xx}$, the characteristic relaxation time of the flow velocity $\tau$ is of the order of the momentum diffusion time by the distance $d \sim \sqrt{\ln A/n_D}$, i.e. $\tau \sim d^2/\nu$ (we recall that the viscosity coefficient $\nu$ also has the meaning of the momentum diffusion coefficient).

## 4. Corrections to self-consistent results due to fluctuations in location of disks

In the previous sections, we have considered values averaged over realizations. Now let us discuss the role of fluctuations in the arrangement of disks, which we will consider random, so that a typical fluctuation of the number of disks in a region containing $N$ disks is $\sqrt{N}$. Let us first discuss in detail the resistance fluctuations in the absence of a magnetic field and then briefly describe the Hall resistance fluctuations.

Within the framework of the effective medium method, we calculated the rate $\tau^{-1}$, "standing on" one of the disks and replacing all other disks with a homogeneous medium with the required relaxation time. It turned out that the result is determined by disks within a circle of radius $R_\varepsilon = R/\varepsilon = \sqrt{v\tau}$. On average, there are $N_\varepsilon = n_D v\tau \sim \ln A$ disks in this circle with a concentration equal to $n_D = N/S$. The typical deviation of the number of disks in this region from its mean value is of the order of $\sqrt{N_\varepsilon} \sim \sqrt{\ln A}$, whence it follows for the concentration fluctuation $\delta n_D \sim \pm n_D / \sqrt{\ln A}$. According to (22) we estimate the inverse time $\tau^{-1}$ fluctuation caused by this density fluctuation as

$$\delta \tau^{-1} \approx \frac{v \delta n_D}{\ln A} + \frac{v \delta n_D}{\ln^2(1/n_D R^2)} \approx \pm \frac{v n_D}{\ln^{3/2} A}.$$

Using now the formula [49]

$$\delta \rho \sim \frac{<(\delta \tau^{-1})^2>}{\tau^{-1}},$$

we obtain for the typical deviation of the resistivity of the sample from its average value the expression

$$\delta \rho_{xx} \sim \frac{\rho_{xx}}{\ln A} \ll \rho_{xx}. \tag{34}$$

This result means, in particular, that in the expression (27) for the mean value of $\rho_{xx}$ the second term in brackets is illegitimate.

In the case of smooth disks, similar estimates of the contribution of fluctuations to the correction to $\Delta \rho_H$ yield:

$$\delta(\Delta \rho_H) \sim \frac{\Delta \rho_H}{\ln A} \ll \Delta \rho_H. \tag{35}$$

In this way, the fluctuations of the value $\Delta \rho_H$ are also small as compared to its mean value.

## 5. Electron fluid velocity in vicinities of disks

In this section we discuss the velocity field and briefly touch on the problem of finding the electric field. Using the first two formulas (2), expression (16) for $\psi$, and the fact that in our problem $\overline{V}_y = 0$, we obtain $\alpha_1 = 0$ and:

$$V_x = \frac{2}{R}\left[\alpha_2 - \frac{\varepsilon\gamma_2 K_0(\varepsilon\rho)}{2} + \left(\gamma_2 S(\rho) - \frac{\alpha_2}{\rho^2}\right)\cos 2\theta + \gamma_1 S(\rho)\sin 2\theta\right],$$

$$V_y = \frac{2}{R}\left[-\frac{\gamma_1\varepsilon K_0(\varepsilon\rho)}{2} - \gamma_1 S(\rho)\cos 2\theta + \left(\gamma_2 S(\rho) - \frac{\alpha_2}{\rho^2}\right)\sin 2\theta\right], \quad S(\rho) = \frac{\varepsilon K_0(\varepsilon\rho)}{2} + \frac{K_1(\varepsilon\rho)}{\rho} - \frac{K_1(\varepsilon)}{\rho^2}.$$

Hence, for the case of rough discs, when $\gamma$ given by the expression (19) and $\gamma_1 = 0$, it follows from the above equations:

$$V_x = \overline{V}_x\left\{-1 + \frac{K_0(\varepsilon\rho)}{K_0(\varepsilon)} + \left[\frac{K_0(\varepsilon\rho)}{K_0(\varepsilon)} - \frac{1}{\rho^2} + \frac{2}{\varepsilon K_0(\varepsilon)}\left(\frac{K_1(\varepsilon\rho)}{\rho} - \frac{K_1(\varepsilon)}{\rho^2}\right)\right]\cos\theta\right\},$$

$$V_y = \overline{V}_x\left[\frac{K_0(\varepsilon\rho)}{K_0(\varepsilon)} - \frac{1}{\rho^2} + \frac{2}{\varepsilon K_0(\varepsilon)}\left(\frac{K_1(\varepsilon\rho)}{\rho} - \frac{K_1(\varepsilon)}{\rho^2}\right)\right]\sin 2\theta.$$

From these expressions in the domain $\varepsilon\rho \ll 1$ we obtain with the logarithmic accuracy:

$$V_x \approx \frac{\overline{V}_x}{\ln(4/\varepsilon^2 \exp 2\gamma_E)}\left[\ln\rho^2 + \left(1 - \frac{1}{\rho^2}\right)\cos 2\theta\right], \quad V_y \approx \frac{\overline{V}_x}{\ln(4/\varepsilon^2 \exp 2\gamma_E)}\left(1 - \frac{1}{\rho^2}\right)\sin 2\theta.$$

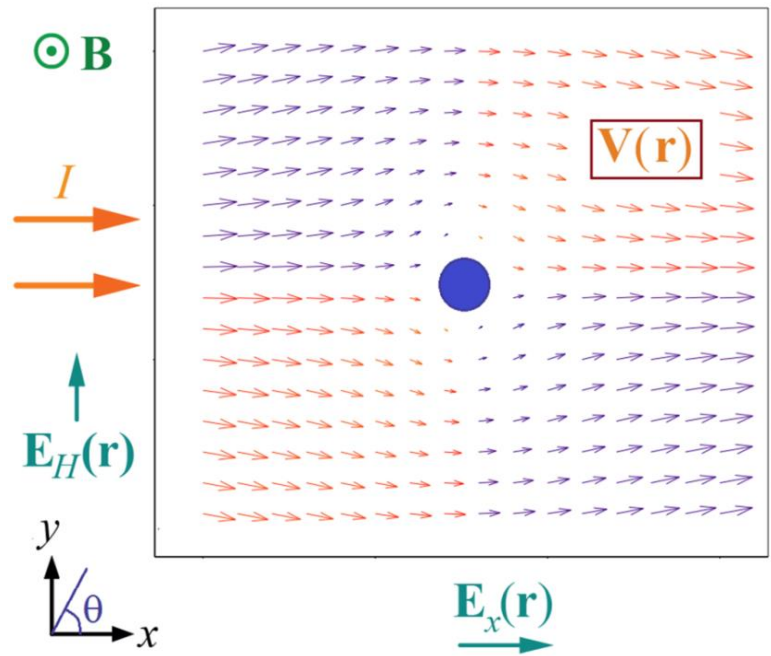

Fig. 2. Profile of the hydrodynamic velocity of the electron fluid near a rough disk.
Size of the plot is 16*R*.

This expressions mean, in particular, that the velocity profile is, on average, symmetrical about an axis passing through the center of a disc and parallel to the abscissa axis.

It is seen from Eq. (24) and the above expressions for the velocity components that, according to the second general statement in Section I, the velocity profile does not depend on the magnetic field [according to Eq. (24) $\varepsilon^2$ is only a function of $n_D R^2$].

Similarly, it is not difficult to find $V_x$ and $V_y$ in the case of smooth disks. The resulting expressions are too cumbersome and we will not give them. Let us note only that the approximation (25) is too crude for $\gamma_2$ and we should use the formula:

$$\gamma_2 \approx \frac{2\alpha_2}{\varepsilon K_0(\varepsilon)} - \frac{2\alpha_2}{4\varepsilon K_0^2(\varepsilon)(1+\beta_2^3)}.$$

Indeed, in the approximation (25) and in the absence of a magnetic field, we get the same expressions for velocity as in the case of rough discs and, therefore, $\mathbf{V}=0$ at the edge of the disk, which is not true. The velocity field now directly depends on the magnetic field and at $B \neq 0$ there is no symmetry with respect to the abscissa axis.

Knowing the velocity profile, we can find the distribution of the electric field everywhere in in the region of the flow from the Navier-Stokes equation for the both the cases of rough and smooth discs.

## III. Beyond standard hydrodynamics

In this section, we present calculations that go beyond hydrodynamics in description of the flow of the electron fluid in different regimes.

There are two reasons for such consideration (both of them were formulated in the Introduction). First, the Navier-Stokes equation is derived from the kinetic equation under the assumption that all harmonics of the distribution function, starting from the third harmonic, relax instantaneously. In fact, the relaxation time of the third harmonic is of the same order as the relaxation time of the second harmonic, or even much longer than it [49]. In the latter case $\tau_4 \approx \tau_2$ and $\tau_5 \approx \tau_3$. According to Appendix C, each of the functions $f_{ns,c}$ is related only to the functions $f_{n\pm 1s,c}$, whence and from the above it follows that the influence of the harmonics $f_{4s,c}, f_{5s,c},$, and so on, on the harmonics $f_{1s,c}, f_{2s,c},$ and $f_{3s,c}$ is rapidly decreasing. The question arises, how the predictions of the theory will change, if the three harmonics are included in the calculations?

Second, it is obvious in advance that there is a thin kinetic layernear the obstacleedges, where many electrons are scattered on sample edged and therefore the hydrodynamic approximation is inapplicable. In this regions, it is necessary to take into account the higher harmonics of the distribution function. Approximation by three harmonics is a step in this direction. We limited ourselves to the case of smooth disks, because in the hydrodynamic approach exactly in this case there appears a correction to the standard Hall resistance, proportional to $v_H$, and it is interesting to see how it will change due to accounting for the third harmonic.

### 1. Basic equations

We start our consideration from the derivation of the exact equations for the momentum flux density tensor $\Pi_{ik}(\mathbf{r})$. Let us write the kinetic equation in the form:

$$\operatorname{div}(f\mathbf{v}) - e\mathbf{E}(\mathbf{r})\frac{\partial f}{\partial \mathbf{p}} + \omega_c \frac{\partial f}{\partial \varphi} = St_{ee}(f), \quad e > 0 \cdot \tag{36}$$

By representing the function $f(\mathbf{p},\mathbf{r})$ in the form of $f = f_0(\varepsilon,\mathbf{r}) + \tilde{f}(\varepsilon,\varphi,\mathbf{r})$, $\int_0^{2\pi} \tilde{f} d\varphi = 0$, multiplying (36) by $p_y$ and by integrating the result with $2d^2\mathbf{p}/(2\pi\hbar)^2$, we obtain:

$$\frac{\partial \bar{\varepsilon}(\mathbf{r})}{\partial y} + en(\mathbf{r})E_y(\mathbf{r}) + n_0 \frac{\partial \Pi_{yx}(\mathbf{r})}{\partial x} + n_0 \frac{\partial \Pi_{yy}(\mathbf{r})}{\partial y} - m\omega_c V_x(\mathbf{r})n(\mathbf{r}) = 0, \tag{37}$$

$$\bar{\varepsilon}(\mathbf{r}) \equiv \int f_0(\varepsilon,\mathbf{r})\varepsilon \frac{2d^2\mathbf{p}}{(2\pi\hbar)^2}, \quad \Pi_{ik}(\mathbf{r}) \equiv \frac{m}{n_0} \int \tilde{f}\, v_i v_k \frac{2d^2\mathbf{p}}{(2\pi\hbar)^2}, \quad V_i(\mathbf{r}) = \frac{1}{n(\mathbf{r})} \int \tilde{f}(\mathbf{p},\mathbf{r}) v_i \frac{2d^2\mathbf{p}}{(2\pi\hbar)^2}$$

where $n(\mathbf{r})$ is the concentration of electrons, $n_0$ is the equilibrium concentration, $\bar{\varepsilon}(\mathbf{r})$ is the energy density of the electron gas, and $\Pi_{ik}(\mathbf{r})$ is viscous stress tensor. We recall that in the case of interparticle collisions there is an equality $\int St_{ee}(f)\mathbf{v}d^2\mathbf{p} = 0$. Similarly, we obtain the equation:

$$\frac{\partial \bar{\varepsilon}(\mathbf{r})}{\partial x} + en(\mathbf{r})E_x(\mathbf{r}) + n_0 \frac{\partial \Pi_{xx}(\mathbf{r})}{\partial x} + n_0 \frac{\partial \Pi_{xy}(\mathbf{r})}{\partial y} + m\omega_c V_y(\mathbf{r})n(\mathbf{r}) = 0 \cdot \tag{38}$$

Note that equations (37) and (38) are exact. They can be conveniently rewritten in the form:

$$enE_x + \frac{\partial \bar{\varepsilon}}{\partial x} + n_0 \operatorname{div} \mathbf{\Pi}_x + mn\omega_c V_y = 0,$$

$$enE_y + \frac{\partial \bar{\varepsilon}}{\partial y} + n_0 \operatorname{div} \mathbf{\Pi}_y - mn\omega_c V_x = 0, \quad \mathbf{\Pi}_i(\mathbf{r}) \equiv \frac{m}{n_0} \int \tilde{f}\, v_i \mathbf{v} \frac{2d^2\mathbf{p}}{(2\pi\hbar)^2} \tag{39}$$

Throughout the paperwe assume that the screening length is smaller than all other spatial scales of the problem, therefore, in absence of current, the electric field is zero. Correspondingly, the concentration and the density of the energy of electrons are the same everywhere.Assuming that the electric field applied to the sample is small, we linearize Eq.(39), that yields:

$$eE_x + \frac{1}{n_0}\frac{\partial \bar{\varepsilon}}{\partial x} + \operatorname{div} \mathbf{\Pi}_x + m\omega_c V_y = 0, \quad eE_y + \frac{1}{n_0}\frac{\partial \bar{\varepsilon}}{\partial y} + \operatorname{div} \mathbf{\Pi}_y - m\omega_c V_x = 0 \cdot \tag{40}$$

Now we introduce the electric $\varphi$ and the electrochemical $\Psi = \varphi - \bar{\varepsilon}/en_0$ potentials. Equations (41) via the last value $\Psi$ are written in the form:

$$\frac{\partial \Psi}{\partial x} = \frac{1}{e}\operatorname{div} \mathbf{\Pi}_x + \frac{B}{c}V_y, \quad \frac{\partial \Psi}{\partial y} = \frac{1}{e}\operatorname{div} \mathbf{\Pi}_y - \frac{B}{c}V_x \cdot \tag{41}$$

By integrating these equations over the flow domain, we obtain the following expressions for the longitudinal and the Hall voltages:

$$U_{sd} \equiv \frac{1}{W}\int_0^W [\Psi(L,y) - \Psi(0,y)]dy = \frac{L}{eS}\sum_j \oint_{\Gamma_j} d\mathbf{l}_j \cdot \mathbf{\Pi}_x + \frac{RL}{S}\sum_j \oint_{\Gamma_j} \Psi(\theta, R)\cos\theta d\theta,$$

$$U_H \equiv \frac{1}{L}\int_0^L [\Psi(x,W) - \Psi(x,0)]dx = \frac{W}{eS}\sum_j \oint_{\Gamma_j} d\mathbf{l}_j \cdot \mathbf{\Pi}_y + \frac{RW}{S}\sum_j \oint_{\Gamma_j} \Psi(\theta, R)\sin\theta d\theta + \frac{BI}{en_0 c},$$
(42)

where $\theta$ is the polar angle vector pointing from the center of this disk to its edge; we assume $\overline{V}_y = 0$ and omit the integrals on the edges of the sample. Equivalent, Eq. (42) can be written via the effective electric field $E_\theta$:

$$U_{sd} = \frac{L}{eS}\sum_j \oint_{\Gamma_j} d\mathbf{l}_j \cdot \mathbf{\Pi}_x + \frac{R^2 L}{S}\sum_j \oint_{\Gamma_j} E_\theta \sin\theta d\theta,$$

$$U_H = \frac{BI}{en_0 c} + \frac{W}{eS}\sum_j \oint_{\Gamma_j} d\mathbf{l}_j \cdot \mathbf{\Pi}_y - \frac{R^2 W}{S}\sum_j \oint_{\Gamma_j} E_\theta \cos\theta d\theta$$
(43)

Here the symbol $E_\theta$ denotes the value $-\partial\Psi/R\partial\theta$, and $d\mathbf{l}_j$ is the vector of the disk boundary element pointing to its center. The first terms in the right-hand sides of Eqs. (43) are equal to the voltage drop on the flow region, and the last terms are equal to the voltage drop on the disks.

Let us transform the second of Eqs. (43). Since $d\mathbf{l}_j \cdot \mathbf{\Pi}_y = -\Pi_{yr} R d\theta$ and $v_y = v_r \sin\theta + v_\theta \cos\theta$, the contribution of the flow region will be written as

$$\frac{W}{eS}\sum_j \oint_{\Gamma_j} d\mathbf{l}_j \cdot \mathbf{\Pi}_y = -\frac{RW}{eS}\sum_j \oint_{\Gamma_j}(\Pi_{rr}\sin\theta + \Pi_{\theta r}\cos\theta)d\theta.$$
(44)

We take into account that, due to the condition $\int_0^{2\pi} \tilde{f}(\varepsilon, \varphi, \mathbf{r})d\varphi = 0$, the relations $\Pi_{yy} = -\Pi_{xx}$, $\Pi_{\theta\theta} = -\Pi_{rr}$ are fulfilled, and the equalities, $V_x \cos\theta + V_y \sin\theta = V_r$, $\partial\Psi/\partial\theta = R\cos\theta\,\partial\Psi/\partial y - R\sin\theta\,\partial\Psi/\partial x$, and $V_r(\mathbf{r})|_{r=R} = 0$ are valid. After cumbersome, but elementary calculations we obtain from these relations and Eq. (41):

$$E_\theta = -\frac{1}{e}\left(\frac{\partial \Pi_{r\theta}}{\partial r} + \frac{2}{r}\Pi_{r\theta} - \frac{1}{r}\frac{\partial \Pi_{rr}}{\partial \theta}\right).$$
(45)

Substituting Eqs. (45) and (44) in formula (43), we obtain for the Hall voltage:

$$U_H = \frac{BI}{en_0 c} + \frac{RW}{eS}\sum_j \oint_{\Gamma_j}\left(\frac{\partial \Pi_{r\theta}}{\partial \rho} - 2\frac{\partial \Pi_{rr}}{\partial \theta}\right)_{\rho=1}\cos\theta d\theta'$$
(46)

or, after averaging over the positions of the disks:

$$U_H = \frac{BI}{en_0 c} + \frac{RWn_D}{e}\oint_{\Gamma}\left(\frac{\partial \Pi_{r\theta}}{\partial \rho} - 2\frac{\partial \Pi_{rr}}{\partial \theta}\right)_{\rho=1}\cos\theta d\theta.$$
(47)

Similarly, for the source-drain voltage we get:

$$U_{sd} = -\frac{RL}{eS} \sum_j \int_{\Gamma_j} \left( \frac{\partial \Pi_{r\theta}}{\partial \rho} - 2\frac{\partial \Pi_{rr}}{\partial \theta} \right)_{\rho=1} \sin\theta \, d\theta \quad \Rightarrow \quad (48)$$

$$U_{sd} = -\frac{RLn_D}{e} \int_\Gamma \left( \frac{\partial \Pi_{r\theta}}{\partial \rho} - 2\frac{\partial \Pi_{rr}}{\partial \theta} \right)_{\rho=1} \sin\theta \, d\theta$$

In the derivation of expressions (47) and (48), it was taken into account that at the edge of the disk the condition $\Pi_{r\theta} = 0$ takes place. We will also need the last expression in the future consideration.

Our approach of description of the ballistic effects is based in the using of a truncated representation of the generalized distribution function $\tilde{f}(\mathbf{r},\varphi)$ as a sum over three angular harmonics:

$$\tilde{f}(\mathbf{r},\varphi) = \sum_{m=1}^{3}[f_{mc}(\mathbf{r})\cos(m\varphi) + f_{ms}(\mathbf{r})\sin(m\varphi)]. \quad (49)$$

Such form of $f$ leads to the following system of the "quasi-hydrodynamic" motion equations of the electron fluid (for their derivation see Appendix E):

$$\frac{\partial \Pi_{xx}}{\partial x} + \frac{\partial \Pi_{xy}}{\partial y} + eE_x + m\omega_c V_y = 0, \quad \frac{\partial \Pi_{yx}}{\partial x} + \frac{\partial \Pi_{yy}}{\partial y} + eE_y - m\omega_c V_x = 0,$$

$$\frac{\partial V_x}{\partial x} + \frac{\partial V_y}{\partial y} = 0, \quad \Pi_{yy} = -\Pi_{xx}, \Pi_{yx} = \Pi_{xy}, \quad (50)$$

$$\Pi_{yx} = -m\nu\left(\frac{\partial V_x}{\partial y} + \frac{\partial V_y}{\partial x}\right) - m\nu_H\left(\frac{\partial V_x}{\partial x} - \frac{\partial V_y}{\partial y}\right) + \frac{\nu\tau_3}{1+\beta_3^2}(\Delta\Pi_{xy} + \beta_3\Delta\Pi_{xx}) + \frac{\nu_H\tau_3}{1+\beta_3^2}(\Delta\Pi_{xx} - \beta_3\Delta\Pi_{xy}),$$

$$\Pi_{xx} = -m\nu\left(\frac{\partial V_x}{\partial x} - \frac{\partial V_y}{\partial y}\right) + m\nu_H\left(\frac{\partial V_x}{\partial y} + \frac{\partial V_y}{\partial x}\right) - \frac{\nu_H\tau_3}{1+\beta_3^2}(\Delta\Pi_{xy} + \beta_3\Delta\Pi_{xx}) + \frac{\nu\tau_3}{1+\beta_3^2}(\Delta\Pi_{xx} - \beta_3\Delta\Pi_{xy})$$

It is impossible to solve these equations in a general form, however it can be done in two special cases, namely, in the case of the Poiseuille flow in a long sample with rough edges and in the case of the flow in a sample with many arbitrary distributed disks. In both these cases the variables describing the flow can be separated, as they were in ordinary hydrodynamics. The first of these problems is considered in Ref. [46]. The second one will be discussed below in the framework of the effective medium method.

In this method, instead of the first two equations (50), one should use the analogous equation with the introduced effective relaxation time $\tau$ and the cyclotron frequency shift $\delta\omega_c$, as it was done in the case of the hydrodynamic equation (8):

$$\frac{\partial \Pi_{xx}}{\partial x} + \frac{\partial \Pi_{xy}}{\partial y} + eE_x + m(\omega_c + \delta\omega_c)V_y = -\frac{mV_x}{\tau},$$
$$\frac{\partial \Pi_{yx}}{\partial x} + \frac{\partial \Pi_{yy}}{\partial y} + eE_y - m(\omega_c + \delta\omega_c)V_x = -\frac{mV_y}{\tau} \quad (51)$$

or, equivalently:

$$\frac{\partial \Pi_{xx}}{\partial x}+\frac{\partial \Pi_{xy}}{\partial y}-\frac{\partial \Phi}{\partial x}=\frac{m}{\tau}\frac{\partial \psi}{\partial y}, \quad \frac{\partial \Pi_{yx}}{\partial x}+\frac{\partial \Pi_{yy}}{\partial y}-\frac{\partial \Phi}{\partial y}=-\frac{m}{\tau}\frac{\partial \psi}{\partial x}. \quad (52)$$

$$\Phi = e\Psi - m(\omega_c + \delta\omega_c)\psi$$

Applying of the rotor operator to equations (52) yields:

$$2\frac{\partial^2 \Pi_{xx}}{\partial y \partial x}+\frac{\partial^2 \Pi_{xy}}{\partial^2 y}-\frac{\partial^2 \Pi_{xy}}{\partial^2 x}=\frac{m}{\tau}\Delta\psi.$$

By substituting here expressions (50) for $\Pi_{ik}$ and using Eqs. (52), we obtain:

$$\left[m\nu+\frac{m\xi^2}{\tau}(1-\beta_2\beta_3)\right]\Delta^2\psi-\xi^2(\beta_2+\beta_3)\Delta^2\Phi=\frac{m}{\tau}\Delta\psi, \quad \xi=\sqrt{\frac{\nu\tau_3}{(1+\beta_3^2)}}. \quad (53)$$

Applying the divergence operator to equations (51b) and performing similar calculations, we obtain another equation for the functions $\Phi$ and $\psi$:

$$m\beta_2\nu\Delta^2\psi+\xi^2(1-\beta_2\beta_3)\Delta^2\Phi-\Delta\Phi=0. \quad (54)$$

Equations (53) and (54) can be written as:

$$\frac{(\beta_2+\beta_3)}{(1+\beta_2^2)m\nu}\Delta\Phi=\left[1+\frac{\lambda^2(1-\beta_2\beta_3)^2}{\lambda_\tau^2(1+\beta_2^2)^2}\right]\Delta^2\psi-\frac{1-\beta_2\beta_3}{\lambda_\tau^2(1+\beta_2^2)}\Delta\psi,$$

$$\frac{m\beta_2}{\tau}\Delta\psi=-\lambda^2\left[1+\frac{\lambda^2(1-\beta_2\beta_3)^2}{\lambda_\tau^2(1+\beta_2^2)^2}\right]\Delta^2\Phi+\left[1+\frac{\lambda^2}{\lambda_\tau^2}\frac{1-\beta_2\beta_3}{1+\beta_2^2}\right]\Delta\Phi, \quad (55)$$

where

$$\lambda^2=\xi^2(1+\beta_2^2)=\frac{\nu_0\tau_3}{1+\beta_3^2}=\frac{l_2 l_3}{4(1+\beta_3^2)}, \quad \nu_0=\nu(0), \quad \lambda_\tau^2=\nu\tau.$$

Expressing now $\Delta\Phi$ from the first equation and substituting in the second, as well as expressing from the second equation $\Delta\psi$ and substituting in the first, we get we get two identical equations:

$$\lambda^2\left[1+\frac{\lambda^2}{\lambda_\tau^2}\frac{(1-\beta_2\beta_3)^2}{(1+\beta_2^2)^2}\right]\Delta^3\psi-\left[1+\frac{2\lambda^2}{\lambda_\tau^2}\frac{1-\beta_2\beta_3}{1+\beta_2^2}\right]\Delta^2\psi+\frac{1}{\lambda_\tau^2}\Delta\psi=0,$$

$$\lambda^2\left[1+\frac{\lambda^2}{\lambda_\tau^2}\frac{(1-\beta_2\beta_3)^2}{(1+\beta_2^2)^2}\right]\Delta^3\Phi-\left[1+\frac{2\lambda^2}{\lambda_\tau^2}\frac{1-\beta_2\beta_3}{1+\beta_2^2}\right]\Delta^2\Phi+\frac{1}{\lambda_\tau^2}\Delta\Phi=0 \quad (56)$$

We conclude this subsection by deriving the boundary conditions at the edge of a smooth disk. The mirror-like reflection of electrons from the disk boundary means that the distribution function is symmetric with respect to the reflection operation with respect to the tangent to this boundary:

$$\tilde{f}(\varepsilon,\varphi,R,\theta)=\tilde{f}(\varepsilon,\pi+2\theta-\varphi,R,\theta). \quad (57)$$

From this condition for *f* we can obtain the two following conditions for $\Pi$ (they are derived in Appendix F):

$$\Pi_{r\theta}|_{r=R} = 0, \quad \left(-\frac{\partial \Pi_{rr}}{\partial r} + \frac{2\Pi_{rr}}{r} + \frac{1}{r}\frac{\partial \Pi_{r\theta}}{\partial \theta}\right)_{r=R} + \beta_3 \left(\frac{\partial \Pi_{r\theta}}{\partial r} + \frac{1}{r}\frac{\partial \Pi_{rr}}{\partial \theta} - \frac{2\Pi_{r\theta}}{r}\right)_{r=R} = 0. \quad (58)$$

The first of these conditions is accurate, while the second, as can be seen from Appendix D, is approximate: it does not take into account the contribution of the fourth harmonic.

## 2. Solution of equations for $\psi$ and $\Phi$ around one disk

Equations (56) can be solved by presenting the differential operator as a product of three commuting differential operators as follows:

$$\Delta(\Delta - \kappa_+)(\Delta - \kappa_-)\psi = 0, \quad (59)$$

where $\kappa_\pm$ are roots of the equation:

$$\lambda^2 \left[1 + \frac{\lambda^2}{\lambda_\tau^2}\frac{(1-\beta_2\beta_3)^2}{(1+\beta_2^2)^2}\right]\kappa^2 - \left[1 + \frac{2\lambda^2}{\lambda_\tau^2}\frac{1-\beta_2\beta_3}{1+\beta_2^2}\right]\kappa + \frac{1}{\lambda_\tau^2} = 0. \quad (60)$$

With corrections of the order of $\lambda^2/\lambda_\tau^2$, the values $\kappa_\pm$ have the form: $\kappa_+ = 1/\lambda^2$ and $\kappa_- = 1/\lambda_\tau^2$. Bearing in mind that we are interested in the first-order harmonic of the function $\psi$, the operator in the left part of Eq. (59) can be written as the product of three radial differential operators:

$$\Delta_r(\Delta_r - \kappa_+)(\Delta_r - \kappa_-)\chi = 0, \quad (61)$$

where $\chi$ is the absolute value of $\psi$, $\psi = 2\operatorname{Re}[\chi \exp(i\theta)]$, and

$$\Delta_r = \frac{1}{r}\frac{d}{dr}\left(r\frac{d}{dr}\right) - \frac{1}{r^2}$$

is the radial part of the Laplace operator. The solution of equation (61) is the sum of the solutions of the three differential equations

$$\Delta_r \chi = 0, \quad (\Delta_r - \kappa_+)\chi = 0, \quad (\Delta_r - \kappa_-)\chi = 0.$$

Such sum can be presented as:

$$\chi = \alpha r + \frac{\delta}{r} + \gamma_+ K_1(\sqrt{\kappa_+}r) + \gamma_- K_1(\sqrt{\kappa_-}r) \approx \alpha r + \frac{\delta}{r} + \gamma_+ K_1(r/\lambda) + \gamma_- K_1(r/\lambda_\tau). \quad (62)$$

Similarly, putting $\Phi = 2\operatorname{Re}[\phi \exp(i\theta)]$, we have:

$$\phi = \tilde{\alpha} r + \frac{\tilde{\delta}}{r} + \tilde{\gamma}_+ K_1(\sqrt{\kappa_+}r) + \tilde{\gamma}_- K_1(\sqrt{\kappa_-}r) \approx \tilde{\alpha} r + \frac{\tilde{\delta}}{r} + \tilde{\gamma}_+ K_1(r/\lambda) + \tilde{\gamma}_- K_1(r/\lambda_\tau). \quad (63)$$

In expressions (62) and (63) we omit the exponentially increasing terms.

By virtue of Eqs. (55), the coefficients in front of the cylindrical functions are not independent. Substituting into the first equation (55) the functions $\tilde{\gamma}_- K_1(\sqrt{\kappa_-}r)$, $\gamma_- K_1(\sqrt{\kappa_-}r)$ and taking into account that $\Delta K_1(\sqrt{\kappa_-}r) = \kappa_- K_1(\sqrt{\kappa_-}r)$, we obtain:

$$\frac{(\beta_2 + \beta_3)}{(1+\beta_2^2)m\nu}\tilde{\gamma}_- = \left\{\left[1 + \frac{\lambda^2(1-\beta_2\beta_3)^2}{\lambda_\tau^2(1+\beta_2^2)^2}\right]\kappa_- - \frac{(1-\beta_2\beta_3)}{\lambda_\tau^2(1+\beta_2^2)}\right\}\gamma_-.$$

Substituting here the root of equation (60), we will have:

$$\tilde{\gamma}_- = \frac{2m\beta_2}{\tau\left[1 + \sqrt{1 - 4\frac{\lambda^2\beta_2(\beta_2+\beta_3)}{\lambda_\tau^2(1+\beta_2^2)^2}}\right]}\gamma_- \approx \frac{m\beta_2}{\tau}\gamma_-. \tag{64}$$

Similarly, using the function $K_1(\sqrt{\kappa_+}r)$, from the second equation (55) we find:

$$\gamma_+ = \frac{2\tau(\beta_2+\beta_3)\lambda^2}{m(1+\beta_2^2)\left[1+\sqrt{1-4\frac{\lambda^2\beta_2(\beta_2+\beta_3)}{\lambda_\tau^2(1+\beta_2^2)^2}}\right]\lambda_\tau^2}\tilde{\gamma}_+ \approx \frac{\tau(\beta_2+\beta_3)\lambda^2}{m(1+\beta_2^2)\lambda_\tau^2}\tilde{\gamma}_+ = \frac{\tau_3(\beta_2+\beta_3)}{m(1+\beta_3^2)}\tilde{\gamma}_+. \tag{65}$$

We see that there are only six of the eight constants introduced above are independent. Indeed, these constants arises from solutions (62) and (63) of equations (56), which were derived from equations (51) by differentiations, that leads to an increase in the number solutions.

Let us now find a solution to equations (52). In radial variables [see formulas (F3)] they are written in the form:

$$\frac{\partial \Pi_{rr}}{\partial r} + \frac{2\Pi_{rr}}{r} + \frac{1}{r}\frac{\partial \Pi_{r\theta}}{\partial \theta} = \frac{\partial \Phi}{\partial r} + \frac{m}{\tau}\frac{1}{r}\frac{\partial \psi}{\partial \theta}, \quad \frac{\partial \Pi_{r\theta}}{\partial r} + \frac{2\Pi_{r\theta}}{r} - \frac{1}{r}\frac{\partial \Pi_{rr}}{\partial \theta} = \frac{1}{r}\frac{\partial \Phi}{\partial \theta} - \frac{m}{\tau}\frac{\partial \psi}{\partial r}. \tag{66}$$

By substituting in them

$$\Phi(\mathbf{r}) = 2\operatorname{Re}\{\phi(r)\exp(i\theta)\}, \quad \Pi_{\alpha\beta}(\mathbf{r}) = 2\operatorname{Re}\{Q_{\alpha\beta}(r)\exp(i\theta)\}, \quad \psi(\mathbf{r}) = 2\operatorname{Re}[\chi\exp(i\theta)],$$

we get:

$$\frac{\partial Q_{rr}}{\partial r} + \frac{2Q_{rr}}{r} + i\frac{Q_{r\theta}}{r} = \frac{\partial \phi}{\partial r} + i\frac{m}{\tau}\frac{\chi}{r}, \quad \frac{\partial Q_{r\theta}}{\partial r} + \frac{2Q_{r\theta}}{r} - i\frac{Q_{rr}}{r} = i\frac{\phi}{r} - \frac{m}{\tau}\frac{\partial \chi}{\partial r}, \tag{67a}$$

or:

$$r^2\frac{\partial^2 Q_{r\theta}}{\partial r^2} + 5r\frac{\partial Q_{r\theta}}{\partial r} + 3Q_{r\theta} = \frac{\partial F}{\partial r}, \quad Q_{rr} = -ir\frac{\partial Q_{r\theta}}{\partial r} - 2iQ_{r\theta} - \phi - \frac{im}{\tau}r\frac{\partial \chi}{\partial r}, \quad F(r) = 2ir\phi - \frac{m}{\tau}r\frac{\partial(r\chi)}{\partial r}. \tag{67b}$$

The following calculations, due to their cumbersome nature, we place in the Appendix G. As a result we obtain the close system of equation for the constants $\gamma_\pm$, $\tilde{\gamma}_\pm$, and the corresponding solutions for $\phi$, $\chi$ and $\Pi_{ik}$.

## 3. Resistivity tensor

First of all, note that, according to expressions (62), (63) and (IE), there are two scales of change of physical quantities as functions of the distance from the center of the disk. They are the characteristic lengths related with the eigenvalues of our problem [which are the roots equation (60)]: $1/\sqrt{\kappa_+} \approx \lambda$ and $1/\sqrt{\kappa_-} \approx \lambda_\tau$.

The scale $\lambda_\tau$ is always much larger than the disk radius and even the distance between the disks, while $\lambda$ can be both much smaller and much larger than the disk radius. In the first case, we have a narrow kinetic layer around the disk in which, strictly speaking, we should solve the kinetic equation, while outside this layer, the standard hydrodynamics "works" and all spatial dependences are slow. In the second case the flow also allows a macroscopic description, although not everywhere in terms of standard hydrodynamics.

Substituting in (47) and (48) the expressions for the components of the tensor $\Pi_{\alpha\beta}(\mathbf{r})$ [see Appendix G, expressions (G3)], after some calculations we obtain:

$$U_{sd} = \frac{2\pi vmn_D L\varepsilon^2}{eR}[\gamma_{-2}\varepsilon_- K_2(\varepsilon_-) + \gamma_{+2}\varepsilon_+ K_2(\varepsilon_+)] \tag{68}$$

and

$$U_H = \frac{BIW}{en_0 c} + \frac{2\pi vmn_D W\varepsilon^2}{eR}[\gamma_{-1}\varepsilon_- K_2(\varepsilon_-) + \gamma_{+1}\varepsilon_+ K_2(\varepsilon_+)], \tag{69}$$

where we introduce the notations: $\varepsilon^2 = R^2/v\tau \ll 1$, $\varepsilon_- = \sqrt{\kappa_-}R \ll 1$, $\varepsilon_+ = \sqrt{\kappa_+}R$, and $\gamma_{\mp 1,2}$ being the real and imaginary parts of the coefficients of $\gamma_{\mp}$.

In our calculations, the inequality $\varepsilon_- \ll 1$ is always satisfied, so we can assume here with sufficient accuracy that $K_2(\varepsilon_-) \approx 2/\varepsilon_-^2$, whereas $\varepsilon_+$ can be any: $\varepsilon_+ \gg 1$ at $\lambda \ll R$ as well as $\varepsilon_+ \ll 1$ at $\lambda \gg R$. In the first case, in spite of the exponential smallness $K_2(\varepsilon_+)$, the second terms in square brackets in expressions (68) and (69) cannot be neglected, as the coefficients $\gamma_{+1,2}$ are exponentially large. Indeed, it can be seen from equations (G8) - (G9) that they include only combinations of $\gamma_+ K_p(\varepsilon_+)$, $p = 0,1,2$.

Let us, first, consider the case of $\lambda \ll R$ in the limit of small magnetic fields. Finding the coefficients $\gamma_\pm$ from the equations (E8)-(E9) and substituting them into Eqs. (68) and (69), we obtain:

$$\rho_{xx} \approx \frac{8\pi vmn_D}{e^2 n_0 \ln A} - \frac{112\pi vmn_D}{e^2 n_0 \ln^2 A}\frac{\lambda^2}{R^2}, \tag{70}$$

$$\Delta \rho_H \equiv \rho_H - \rho_H^0 = \frac{8\pi v m n_D}{e^2 n_0 \ln^2 A} \beta_2 + \frac{16\pi v m n_D}{e^2 n_0 \ln^2 A} \frac{\lambda^2}{R^2}(8\beta_2 - \beta_3). \tag{71}$$

In both expressions, the main contribution given by the first termsthat coincide with the hydrodynamic results. As for the second terms, for the diagonal resistivity $\rho_{xx}$ it is small, while the situation for $\Delta \rho_H$ is more complicated.

The magnitude and sign of the second term in Eq. (71) depend on the relation between the lengths $l_2$, $l_3$ and $R$. If relaxation lengths are of the same order, the second terms are small comparedto the first ones, so that the hydrodynamic corrections dominate. This demonstrates the stability of hydrodynamics, the equations of which were derived under the assumption $l_3 \ll l_2$. More interesting is the situation when $l_3 \gg l_2$, which is realized in a degenerate electron Fermi gas [49]. In this case and under the additional condition $l_3 \gg R$ the second term in the right part of (71) is larger than the first term and has a different sign. Thus we have:

$$\Delta \rho_H \approx -\frac{16\pi v m n_D}{e^2 n_0 \ln^2 A} \frac{\lambda^2}{R^2} \beta_3 = -\frac{6\pi v m n_D}{e^2 n_0 \ln^2 A} \frac{l_3^2}{R^2} \beta_2. \tag{72}$$

In the case $\lambda \gg R$ it turns out:

$$\rho_{xx} = \frac{20\pi v m n_D R^2}{e^2 n_0 [26\lambda^2 + 5R^2 \ln A^{1/2}]}, \tag{73}$$

$$\Delta \rho_H = -\frac{88\pi v m n_D R^2 \lambda^2}{e^2 n_0 [26\lambda^2 + 5R^2 \ln A^{1/2}]^2} \beta_3. \tag{74}$$

At the conditions $R^2 \ll \lambda^2 \ll R^2 \ln A^{1/2}$ the main contributions to (70) and (73) coincide, and (74) differs from (72) only by a factor close to unity. At $\lambda^2 \gg R^2 \ln(\lambda_\tau / R)$ we arrive to unexpected results:

$$\rho_{xx} \approx \frac{10\pi v m n_D R^2}{13 e^2 n_0 \lambda^2} = \frac{10\pi m n_D R^2}{13 e^2 n_0 \tau_3}, \tag{75}$$

$$\Delta \rho_H \approx -\frac{88\pi v m n_D R^2}{676 e^2 n_0 \lambda^2} \beta_3 = -\frac{66\pi R^2 n_D}{169} \rho_H^0. \tag{76}$$

It is noteworthy that that the longitudinal resistance does not depend on the relaxation time of the second harmonic of the distribution function, i.e., on the viscous stress relaxation time, and the correction to the Hall resistance does not depend on the relaxation times of both the second and third harmonics, but depends only on the size and concentration of disks! This means that within the three-harmonic approximation under consideration, there is a regime that is radically different from the hydrodynamic one. From (75) we see that $\tau \sim \tau_3 / n_D R^2$, whence it follows $\lambda_\tau \sim \lambda / \sqrt{n_D R^2} \gg \lambda$.

The expression (73) can be interpreted as the resistance of the system of two parallel channels, the hydrodynamic one and the non-hydrodynamic one:

$$\frac{1}{\rho_{xx}} = \frac{e^2 n_0 \tau_h}{m} + \frac{e^2 n_0 \tau_{nh}}{m}, \quad \tau_h = \frac{\ln A}{2\pi l_2^2 n_D} \tau_2 \gg \tau_2, \quad \tau_{nh} = \frac{13}{10\pi n_D R^2} \tau_3 \gg \tau_3.$$

The first term in the right-hand side of this expression coincides with the conductivity in the hydrodynamic regime, and the second term is the conductivity in the non-hydrodynamic regime discussed above. Introducing notation $\Pi_{ik,h}$ for the hydrodynamic viscosity tensor

$$\Pi_{xx,h} = -\Pi_{yy,h} = -m\nu \left( \frac{\partial V_x}{\partial x} - \frac{\partial V_y}{\partial y} \right), \quad \Pi_{xy,h} = \Pi_{yx,h} = -m\nu \left( \frac{\partial V_x}{\partial y} + \frac{\partial V_y}{\partial x} \right),$$

and using Eq.(50) we obtain a diffusion-like equation for dynamics of the shear stress:

$$\frac{1}{\tau_2} \Pi_{ik} = \frac{1}{\tau_2} \Pi_{ik,h} + \nu_3 \Delta \Pi_{ik}, \quad \nu_3 = \frac{1}{4} v_F l_3.$$

In the hydrodynamic regime, the first term dominates in the right part of this equation, whereas in the non-hydrodynamic regime, when $R \ll \lambda$, the second term plays the main role in the regions $r < \lambda$. In this case, the balance between relaxation and diffusion of the shear stress tensor is established:

$$-\nu_3 \Delta \Pi_{ik} = -\frac{1}{\tau_2} \Pi_{ik}.$$

The voltage applied to the sample is approximately equal to the sum of the voltages on regions $r < \lambda$ and on the internal areas of disk.

Here we note that the described picture breaks down already in very small magnetic fields when the inequality $\lambda < R$, equivalent to the inequality

$$\beta_2 > \frac{l_2}{3R} \sqrt{\frac{\tau_2}{\tau_3}} \ll 1,$$

becomes valid.

It is instructive to compare these results with the results of calculation of the Poiseuille flow in a long sample within the three-harmonic approximation. The last consideration was performed in recent work [46]. For such flow the Hall voltage takes the form:

$$eU_H = \frac{B}{n_0 c} I - \Pi_{xx}(W) - \Pi_{xx}(0). \tag{77}$$

In narrow samples, $l_2, l_3 \ll W$, this general formula yields:

$$U_H = \frac{B}{en_0 c} I - \beta_2 E_x W + \frac{\beta}{1+\lambda_3/\lambda} E_x W, \quad \lambda_3/\lambda \sim \frac{1}{\sqrt{1+\beta_3^2}} \sqrt{\frac{l_3}{l_2}}. \tag{78}$$

The first two terms in the right-hand side of this expression are the hydrodynamic bulk contribution: the first one is the main part, whereas the second one is the Hall viscosity correction. The third term in Eq. (78) corresponds to the voltage drop in the narrow, with the width of order of $\sqrt{l_3 l_2}/(1+\beta_3^2) \ll W$, near-edge layers.

We see that the hydrodynamic correction by its absolute value is always larger than the ballistic correction due to the third harmonic. It is also seen that at $l_3 \gg l_2$ the hydrodynamic correction dominates, while at $l_3 \approx l_2$ both the corrections are of the same order of magnitude and partially compensate each other. It is noteworthy that for the flow bypassing disks the situation is opposite: at $l_3 \gg l_2$ the third-harmonic ballistic correction may be the main one, while at $l_3 \approx l_2$ it is small as compared with the hydrodynamic one.

Let us briefly discuss the limit $l_3 \to 0$ (the purely hydrodynamic limit).

In the system with disks in this limit there, as expected, only the hydrodynamic correction remains, while in the case of the Poiseuille flow the corrections compensate each other. In other words, for a Poiseuille problem the result in the limit $l_3 \to 0$ does not coincide with the formulas where we formally put at $l_3 = 0$. It is not difficult to understand how it happens. Indeed, at $l_3 \to 0$ the width of the edge layers $\sim \sqrt{l_2 l_3}$ tends to zero, while the derivative of velocity and other quantities tends to infinity. Correspondingly, at the edges "jumps" or boundary layers are formed, where the function $\Pi_{xx}(y)$, whose values at edges determine the Hall voltage according to Eq. (78), sharply increases. Thus, all quantities tend to their hydrodynamic values at all points except the very sample edges. Such singularity is obtained in the framework of the accepted model, however, it is not excluded that taking into account higher moments in the real problem may somehow "cure" it.

In this way, we see that the answer to the question about the magnitude of the correction to the Hall resistance due to the third harmonic is not obvious in advance and depends on the system under study.

## IV. Comparison with experiment

### 1. Qualitative comparison

**1.1. General similarity of experimental data and predictions of theory.** Let us compare the theoretical results on longitudinal resistivity obtained in Section II with the experimental results of Ref. [27]. As noted in Introduction, in this work a set of high-quality GaAs quantum well samples were fabricated in which localized macroscopic obstacles of various densities were made using electron-beam lithography and subsequent reactive ion etching. Extensive magnetotransport measurements of these samples were performed in [27], and various types of the giant negative magnetoresistance were observed.

In Fig 3(a-c) we present an electron microscope photography of a typical sample studied in experiment [27] and results on the longitudinal resistance in moderate magnetic fields at different temperatures. It is seen that a strong negative magnetoresistance with some additional features, depending on temperature, was observed. In Fig. 3(d) we present the result of our calculation of longitudinal resistance in arbitrary units for the samples with rough and smooth discs at two different temperatures. It is seen that the experimental and the theoretical curves are qualitatively correspond one to another very well.

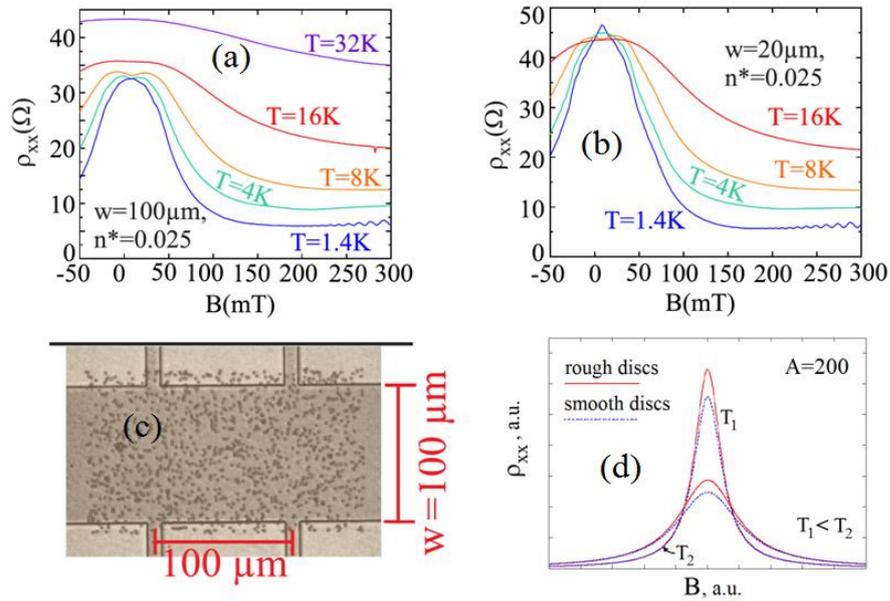

Fig. 3. (a,b) Magnetoresistance of GaAs quantum well samples with obstacles fabricated by ion etching at different temperatures $T$ reported in [27]. The samples the data for which are presented in panels (a) and (b) differ by their widths $w$, but have the same dimensionless densities of obstacles, $n^*=n_D R^2$. The obstacle radius $R$ is 0.42 μm. (c) Scanning electron microscope photography of a typical sample. Panels (a-c) are taken from Ref. [27]. (d) The results of the developed theory for two different temperatures, $T_1$, $T_2$, $T_2=1.5T_1$ for the cases of rough and smooth discs [equations (23) and (27)].

**1.2. Comparison of shapers of experimental and theoretical magnetoresistance.** Moreover, we compared the shape of the calculated magnetoresistance curves for the samples with rough and smooth disks studied theoretically with the shape of the experimental magnetoresistance curve corresponding to the lowest temperature (see Fig. 4). It is seen that the calculation result for the disks with smooth edges, from which the electrons were reflected mirror-like, agrees with the experimental curve much better than the calculation result for the disks with rough edges, which scatter electrons diffusely. The theoretical curve for the latter case in the region of small magnetic fields is sharper than both the experimental and theoretical curves for the smooth disks.

Note that we use arbitrary units in Fig.4 for the resistance as in the current subsection we perform only a qualitative comparison of the experimental data with our predictions.

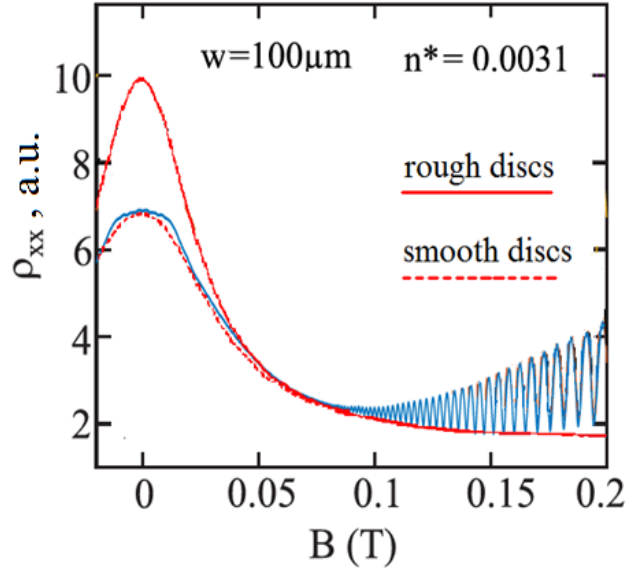

Fig. 4. Magnetoresistance of a GaAs quantum well sample studied in Ref. [27] with obstacles made by ion etching at temperatures *T = 80 mK* . Experimental bluecurve is taken from Ref. [27]. Solid and dashed red curves are the results of our hydrodynamic theoryfor the samples with the rough and the smooth discs. The last curves are plotted by Eqs. (23) and (27), respectively .

Based on results of Fig. 4, we conclude that a detailed comparison of theory and experiment allows to choose the a realistic model of sampleswith macroscopic obstacles.

## 2. Quantitative comparison

**2.1. Half-width of magnetoresistance curves.** Let us qualitatively discuss the magnetoresistance curves for the two samples studied in Ref. [27]for some intermediate temperature, for example, for 8К [see Fig.3(b)].

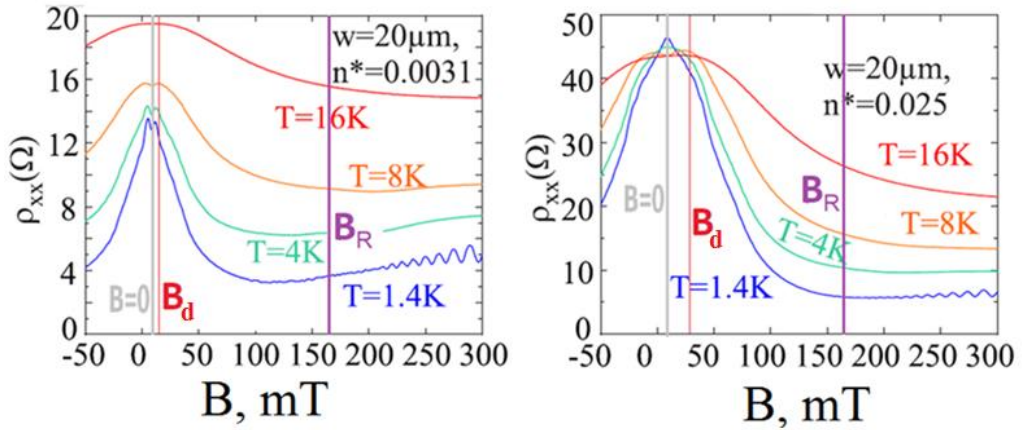

Fig. 5. Panels (a) and (b) from Fig. 4 with added vertical lines correspondingto the characteristic magnetic fields $B_R$ and $B_d$. Vertical grey line depicts the central of the magnetoresistance curve corresponding to zero magnetic field inside the sample (the scale on horizontal axis, apparently, has an artificial shift)

To characterize the width of magnetoresistance curves, we introduce their half-width $B_{1/2}$. Within the hydrodynamic theory [17,49], this value is determined by the relaxation time of the shear stress $\tau_2$ from the condition $\omega_c(B_{1/2})\tau_2 = 1$. In all the samples studied in [27] the radii of discs are R=0.5 мкм. For sample «1», in which the density of defects is $n_D$=1.24*10$^6$ cm$^{-2}$ [the corresponding mean distance between defects is d= $(n_D)^{-1/2}$=9.0 μm], we obtain for the relaxation length the result: $l_2$= $v_F\tau_2$= 1.5 μm from the value $B_{1/2}$ extracted from the magnetoresistance width. For sample «2», in which the defect density is $n_D$= 1.0*10$^7$μm$^{-2}$ [in this sample we have $d$= $(n_D)^{-1/2}$=3.2 μm], we deduce within the same procedure: $l_2$=$v_F\tau_2$=1.1 μm.

Our analysis shows that the mean free path lengths $l_2$ are smaller than the mean distances between the defects in both the samples as the inequality d>$l_2$ is fulfilled. Thus a large hydrodynamic contribution to the flow is expected in both samples.

We emphasize that the parameter **$B_{1/2}$** is the only fitting parameter in our analysis of experimental data on the giant negative magnetoresistance. It yields the value $\tau_2$, which, in its turn, leads the hydrodynamic contribution to the resistance at zero magnetic field $\Delta\rho(B=0)$. Below we compare the values $\Delta\rho(B=0)$ calculated for both the samples with $n_D$=1.24·10$^6$ cm$^{-2}$ and 10·10$^6$ cm$^{-2}$ with the experimental data on the values $\Delta\rho(B=0)$ in these two samples.

**2.2. Important characteristic magnetic fields.** The important characteristic values of the magnetic field are the magnetic field $B_R$, at which the cyclotron radius takes the value equal to the defect radius, and the field $B_d$, at which the cyclotron radius turns out to be of the order of the distance between the defects $d$. For both samples the field $B_R$ is the same, being equal to $164\,\text{mT}$, and the field $B_d$ is equal 9.2 mT for sample "1" and $26.0\,\text{mT}$ for sample "2". Both these fields are shown by vertical lines in Fig. 5.

The physical nature of the regimes determined by these field are as follows. At B<$B_d$ when $R_c$>d the ballistic effects are important, related with scatterings of electrons only on discs, without a substantial role of the interparticle scattering. In the diapason $B_d$<B<$B_R$, corresponding to the inequality R <$R_c$< d, viscous flows are formed in the regions between discs, at the distances greater $R_c$ from disc edges. When B>$B_R$ one should expect a well formed viscous flow everywhere between discs and with some hydrodynamic boundary conditions at the disc edges. The last ones are formed in the semi-ballistic layers around disc edges with the width of the order $R_c$.

**2.3. Amplitude of magnetoresistance.** Now we compare the theoretical and the experimental values of the relative amplitudes of magnetoresistance, which is the difference of the resistance at B=0 and in the limit B>>$B_{1/2}$.

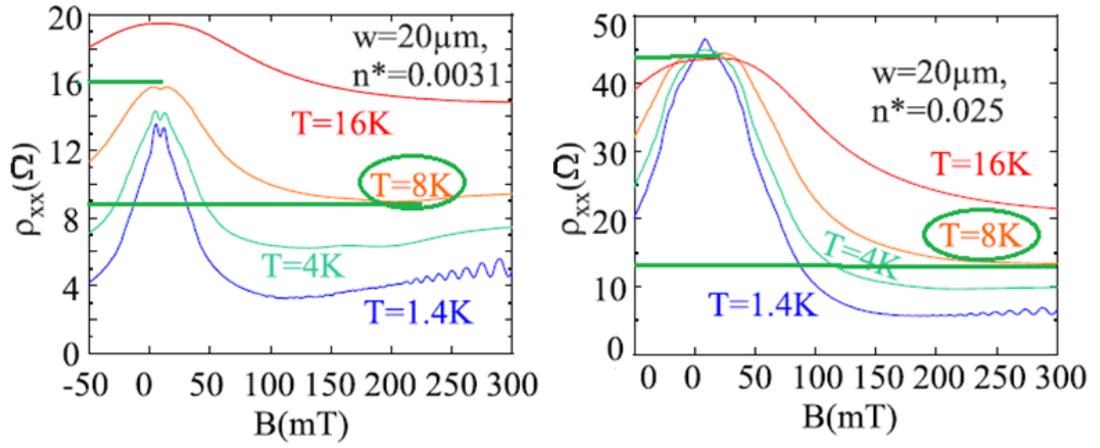

**Fig.6. Panels (a) and (b) from Fig. 4 with added horizontal lines corresponding to the relative height of the magnetoresistance curves at 8 K**

For sample «1» with the defect density $n_D=1.24*10^6$ cm$^{-2}$ (which corresponds to d/R=18), the magnetoresistance amplitude is $\Delta\rho_{exp}(B=0) \sim 7$ Ohm, while the calculation by Eqs.(23) and (27) with the obtained above relaxation time $\tau_2$ yields $\Delta\rho_{theor}(B=0)= 9.6$ Ohm. For sample «2», in which $n_D= 1.0*10^7$cm$^{-2}$ (this value corresponds to d/R=6) we have $\Delta\rho_{exp}(B=0) \sim 35$ Ohm, while equations (23) and (27) without the logarithm factor $\ln(A)$, which is of the order of unity in this case, yield: $\Delta\rho_{theor}(B=0)\sim 142$ Ohm (recall that our calculations assumed the inequality of $\ln(A) \gg 1$).

We see that the theoretical values of the hydrodynamic contribution to the resistance at zero magnetic field, $\Delta\rho(B=0)$, calculated with the tine $\tau_2$ obtained from the widths of the magnetoresistance curves, are in a reasonable agreement with the experimental values. For the discussed two samples, both the values d and R correspond to not too large or even not large (~1) parameter $A=(8\pi n_D R^2)^{-1}$ from Eq. (22), by the logarithm of which, $\ln(A)$, the decomposition of the solution for the flow was performed. Namely, for sample «1» we have $A=12.8$ and $\ln(A) = 2.6$. As a result, a good agreement between the theoretical and the experimental values of $\Delta\rho_{theor}(B)$ is reached. For sample «2» this value turns out to be: $A=1.5$, and thus $\ln(A) = 0.47$. So, in the last case, the developed theory is applicable only on a qualitative, but not quantitative level, the values $\Delta\rho_{theor}(B=0)$ and $\Delta\rho_{exp}(B=0)$ are of the same order of magnitude, but are numerically different.

**2.4. Residual Ohmic resistance.** In the limit of very low temperatures, when the interparticle scattering time $\tau_{ee}\sim 1/T^2$ becomes very long, the giant negative magnetoresistance, does persist (see Fig. 3). The reason of this fact can be in the scattering of electrons on disorder (for example, short-range defects) between the macroscopic defects (discs). Such scattering leads to relaxation of the shear stress with the rate $1/\tau_{2,dis}$ as well as to some residual momentum relaxation with the rate $1/\tau_{1,dis}$ leading to the residual resistance. The last value exhibits itself by the longitudinal resistivity $\rho_{xx}(B)$ in the limit $\omega_c\tau_{2,dis}\gg 1$ [17,50].

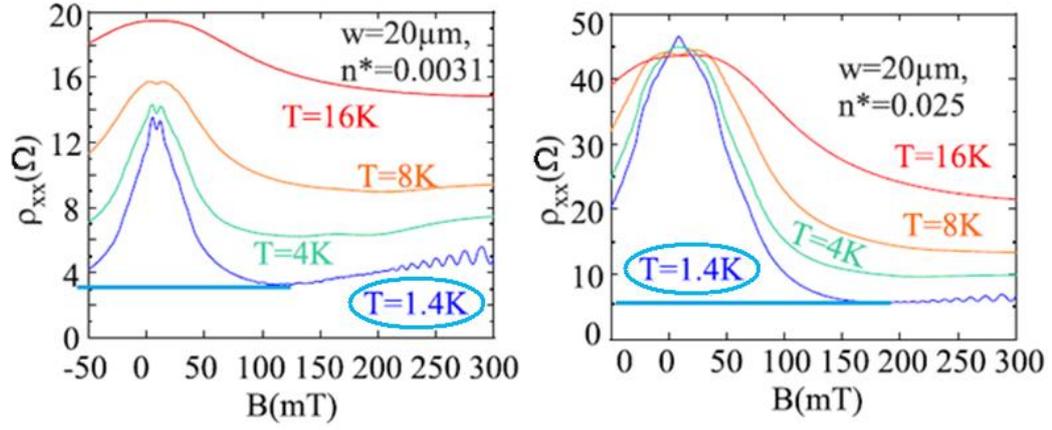

Fig. 7.Panels (a) and (b) from Fig. 4 with added horizontal line correspondingto the resistance in the limit B>> $B_R$ at lowest temperatures 1.4 K

Now we perform a quantitative analysys of experimental data for the case of the lowest temperature T=1.4 K (see Fig. 7). From the the half-width of the curve $\rho_{xx}(B)$ we extract the time $\tau_{2,dis}$ by the procedure described above. From the value$\rho_x$ in the limit $\omega_c\tau_{2,dis}$ >> 1 and the Drude formula$\rho_{xx}$=m/($e^2 n_0 \tau_{1,dis}$) we obtain the experimental value of the transport time $\tau_{1,dis}$. The resulting times$\tau_{1,dis}$и$\tau_{2,dis}$differ one from other anomaleously strongly. For sample «1»we have: $\tau_{1,dis}$ /$\tau_{2,dis}$= 45, and for sample «2» we have:$\tau_{1,dis}$ /$\tau_{2,dis}$= 32. Similar very strong difference of $\tau_{1,dis}$ and $\tau_{2,dis}$ was also found when analyzing experimental data obtained on similar samples (see, for example, [14,17,50].

Ordinary, the values of times $\tau_{1,dis}$, $\tau_{2,dis}$, $\tau_{3,dis}$, etc. have the same order of magnitude. A possible reason for the anomalously large difference between $\tau_{1,dis}$ and $\tau_{2,dis}$, maybe related to the following mechanism. It was proposed in Ref. [52] that dynamically connected pairs of electrons, which repeatedly collide one with other due to their returns induced by the action of magnetic field, play an important role in transport of 2D electrons at classical magnetic fields. Dynamics of such pairs induced the memory effects in ac hydrodynamic magnetotransport, in particular the magnetooscillations of photoconductivity those may be a mechanism of the well-known MIRO effect in high-quality samples. It is noteworthy that similar memory effects in magnetotransport, induced by the analog of such "extended collisions", have been previously studied for the case of non-interacting electrons in disordered samples with localized defects (see, for example, the work [51] and references therein).

For a rigorous description of the effects associated with such extended collisions, leading to a sharp increase in the degree of correlation of the dynamics of electrons, the Boltzmann equation for the one-particle distribution function of electronsis not sufficient. Possibly, it is necessary to solve the equation not only for the evolution of such distribution function, but also for the evolution of spatially inhomogeneous two-particles correlators of electron distributions. For example, the construction of such equations was carried out in Ref. [53] (and in other works of the authors of Ref. [53]) for the Boltzmann gas of uncharged particles and in Ref. [54] for a degenerated gas of electrons.We are currently studying these effects in relation to our problem of a proper explanation of the giant magnetoresistanceof high-quality samples.

## V. Conclusion

We theoretically studied the magnetotransport of electrons in samples with macroscopic defects (disks) in both the hydrodynamic and the quasihydrodynamic regimes. The latter assumes that the third-order harmonics of the electron distribution function as well as the first and the second ones, control the flow, in contrast to the hydrodynamic regime, which is based on the approximation of two harmonics.

We have shown that in the hydrodynamic regime the resistivity tensor does not depend on the Hall viscosity if the boundary conditions on the obstacles do not depend on it, and that the flux profile is independent of the magnetic field if the boundary conditions do not depend on it. We have examined both the case of rough and smooth disc boundaries and have shown that in both cases there is a strong negative magnetoresistance described by the same expression, up to small corrections. As for the Hall resistivity, in the case of rough boundaries of the disks it is exactly equal to the standard value $B/en_0 c$, because of the boundary conditions in this case do not depend on the magnetic field and the Hall viscosity. In the case of smooth boundaries, there is a small correction to the standard value proportional to the Hall viscosity (the boundary conditions depend on the Hall viscosity and the magnetic field).

Based on the obtained results, we argue that the correction to the standard value the Hall resistivity is an important characteristic of the type of the defect edges. Namely, the smother is the disk edges, the larger is the deviation of the Hall resistance from its standard value. Moreover, the dependence of the correction in the Hall resistance on the magnetic field provide information about the relationship between the relaxation times of the second and higher harmonics of the distribution function. However, more precise experiments than [13,27] are needed for this purpose.

In the second part of the paper, we derive "quasihydrodynamic" equations from the kinetic equation based on the three-harmonic approximation for the distribution function and solve them for the case of smooth disks. It is shown that, depending on the relation between the relaxation lengths $l_2, l_3$ and the disk radius $R$, the expression for the resistivity tensor may coincide with the hydrodynamic one or may differ from it. We assume that throughout the article the inequality $l_2 \ll R$ is fulfilled, whereas the length $l_3$ can be any. At $l_3 \ll R$, the resistivity tensor coincides with the hydrodynamic one, regardless of the relation between $l_2$ and $l_3$. In other words, our calculations show that hydrodynamics "works" not only under the condition $l_3 \ll l_2$, which is usually taken when deriving the equations of hydrodynamics from the kinetic equation, but also under any relation between $l_3 и l_2$, if $\sqrt{l_3 l_2} \ll R$ and $l_3 \ll R$. If $\sqrt{l_3 l_2} \ll R$ and $l_3 \gg R$, the longitudinal resistivity still remains hydrodynamic, while the correction to the standard Hall resistivity becomes nonhydrodynamic. Under the condition $\sqrt{l_3 l_2} \gg R$ the situation is different. At a sufficiently large value of $\sqrt{l_3 l_2}$ (the exact inequality is given in the main text of the paper), the flow regime that is as far from the hydrodynamic one is realized. In this regime the longitudinal resistivity is independent of $l_2$, and the correction $\delta\rho_H$ to the standard the Hall resistivity $\rho_H^0$ is independent of both lengths $l_3$ и $l_2$, but depends only on the radius and concentration $n_D$ of the discs, $\delta\rho_H \sim n_D R^2 \rho_H^0$.

We have compared the obtained theoretical results with experiment [27] on the magnetotransport of GaAs quantum well samples with artificially made localized defects of different densities. A good qualitative agreement between the theoretical and experimental magnetoresistance curves is demonstrated. The analysis shows that the model of smooth disks with a relatively high density of disks describes the experimental data better than the model of rough disks. We have also performed a quantitative analysis of the experimental data on the parametersof magnetoresistance curves for two samples with different densities of defects.

We consider that our results will be useful in analyzing and interpreting experiments on hydrodynamic transport in samples with two-dimensional viscous electron fluid and macroscopic localized defects. In particular, we hope that the type of giant negative magnetoresistance, which in [24] is called the bell-shaped magnetoresistance, is explained by our mechanism. This is supported by absence of dependence of the magnetoresistance on the width of the sample and the temperature dependence typical for the magnetoresistance related to the diagonal viscosity (see discussion in [17]).

Based on the obtained results, we also argue that the correction to the standard value the Hall resistivity is an important characteristic of the type of the defect edges. Namely, the smother is the disk edges, the larger is the deviation of the Hall resistance from its standard value. Moreover, the dependence of the correction in the Hall resistance on the magnetic field provide information about the relationship between the relaxation times of the second and higher harmonics of the distribution function.

We thank I. V. Gornyi and D.G. Polyakov, who took an active part in the discussions throughout the work on the problem. They plan to publush soon an article devoted to the same problem, but considered in a slightly different approach. In their work, some other aspects of the problem are planned to be considered with a closer attention.
The work was supported by the Russian Science Foundation (Grant No. 22-12-00139).

## Appendix A
### Expressions for hydrodynamic velocities averaged over flow region

In this section, we demonstrate how do the terms with sample-averaged velocities $\overline{V}_x$ and $\overline{V}_y$ arise in equations (8). For example, the transfer of the function $-m\omega_c\psi/e$ to the right-hand side of (6) in the first equation gives:

$$\frac{m\omega_c}{eS}\int_0^W [\psi(L,y)-\psi(0,y)]dy.$$

Let us now calculate the sample average velocity:

$$\frac{m\omega_c}{e}\overline{V}_y = \frac{m\omega_c}{eS}\int_0^W dy\int_0^L dx\, V_y(x,y) = \frac{m\omega_c}{eS}\int_0^W dy\int_0^L dx\, \frac{\partial\psi(x,y)}{\partial x} = \frac{Rm\omega_c}{eS}\sum_k \oint_{\Gamma_k}\psi\cos\theta_k\, d\theta_k$$
$$+ \frac{m\omega_c}{eS}\int_0^W [\psi(L,y)-\psi(0,y)]dy.$$
(A1)

Here, the second term in the right part coincides with the above expression arising from the transfer of the function $-m\omega_c\psi/e$ to the right part of (6), while the first term is zero because at the edges of the disks $\psi = \mathrm{cons}$ due to the condition $V_r = \partial\psi/r\partial\theta = 0$.

In this way, equation (A1) connect the mean velocity $\overline{V}_y$ with the expressions containing the function $\psi$ along the sample and the disc edges.

# Appendix B
## Generalization of theorem about absence of dependence of resistivity tensor on Hall viscosity

Here we consider sample with macroscopic defects of any form. We will show that properties of the electron fluid is also independent on the Hall viscosity if the defects edges are rough. Thus, the corresponding theorem formulated in Section II.1 has a general character.

In the case of an arbitrarily shaped obstacle, instead of equations (9) we will have the equations:

$$-m\omega_c\overline{V}_y - \frac{eU_{sd}}{L} - \frac{R}{S}\sum_k\oint_{\Gamma_k}\tilde{\Psi}dy - \frac{m\nu R}{eS}\sum_k\oint_{\Gamma_k}\Omega dx = 0,$$
$$m\omega_c\overline{V}_x - \frac{eU_H}{W} + \frac{R}{S}\sum_k\oint_{\Gamma_k}\tilde{\Psi}dx - \frac{m\nu R}{S}\sum_k\oint_{\Gamma_k}\Omega dy = 0.$$
(B1)

From equation (5) it follows that $\Omega$ does not depend on the Hall viscosity, and from equations (4) it follows that the flow function can be presented in the form:

$$\tilde{\Psi} = \tilde{\Psi}_1 + \tilde{\Psi}_2,$$

where $\tilde{\Psi}_1(x,y)$ does not depend on $\nu_H$, but $\tilde{\Psi}_2$ does not depend on the coordinates. Since $\int_{\Gamma_k}\mathrm{const}(x,y)dx = \int_{\Gamma_k}\mathrm{const}(x,y)dy = 0$, it immediately follows from (B1) that the voltages $U_{sd}$ and $U_H$, and, thus, the components of the resistivity tensor are independent of the Hall viscosity. The independence of the velocity profile in the sample with such boundary conditions on the magnetic field follows from equation (5).

# Appendix C
## Calculation of the work of the Hall viscosity term in the Navier-Stokes equation

Below we calculatethe work from the force originated from the Hall viscosity term and show that it is equal to zero:

$$A_H = \frac{m\nu_H}{e}\int [\Delta \mathbf{V} \times \mathbf{e}_z]\mathbf{V}d^2\mathbf{r} = \frac{m\nu_H}{e}\int (V_x \Delta V_y - V_y \Delta V_x)d^2\mathbf{r} = \frac{m\nu_H}{e}\int \text{div}(V_x \nabla V_y - V_y \nabla V_x)d^2\mathbf{r} =$$

$$-\frac{m\nu_H R}{e}\sum_k \oint_{\Gamma_k}\left(V_x \frac{\partial V_y}{\partial r} - V_y \frac{\partial V_x}{\partial r}\right)d\theta = (V_r = 0) =$$

$$\frac{m\nu_H R}{e}\sum_k \oint_{\Gamma_k}\left((V_r \cos\theta - V_\theta \sin\theta)\frac{\partial(V_r \sin\theta + V_\theta \cos\theta)}{\partial r} - (V_r \sin\theta + V_\theta \cos\theta)\frac{\partial(V_r \cos\theta - V_\theta \sin\theta)}{\partial r}\right)d\theta =$$

$$-\frac{m\nu_H R}{e}\sum_k \oint_{\Gamma_k} V_\theta \frac{\partial V_r}{\partial r}d\theta = \left(V_\theta = \frac{\partial \psi}{\partial r}, V_r = -\frac{1}{r}\frac{\partial \psi}{\partial \theta}\right) = \frac{m\nu_H R}{e}\sum_k \oint_{\Gamma_k} \frac{\partial \psi}{\partial r}\frac{\partial}{\partial r}\left(\frac{1}{r}\frac{\partial \psi}{\partial \theta}\right)d\theta =$$

$$-\frac{m\nu_H}{eR}\sum_k \oint_{\Gamma_k} \frac{\partial \psi}{\partial r}\frac{\partial \psi}{\partial \theta}d\theta + \frac{m\nu_H}{e}\sum_k \oint_{\Gamma_k} \frac{\partial \psi}{\partial r}\frac{\partial^2 \psi}{\partial \theta \partial r}d\theta = 0 + \frac{m\nu_H}{2e}\sum_k \oint_{\Gamma_k} \frac{\partial}{\partial \theta}\left(\frac{\partial \psi}{\partial r}\right)^2 d\theta = 0 + 0 = 0$$

# Appendix D
## Derivation of the effective medium equation

Let us start with the derivation of equations (10) for the case of the absence of the magnetic field (its consideration is easy). For this purpose, let us introduce the function

$$H(\mathbf{r};\mathbf{r}_1,...\mathbf{r}_N) = 1 - \sum_{j=1}^{N} h(R - |\mathbf{r} - \mathbf{r}_j|), \tag{D1}$$

where $h(x)$ is the stepped Heaviside function and the points $\mathbf{r}_j$ describe the positions of the centers of the disks. In addition, we introduce the probability density of the given disk configuration $P(\mathbf{r}_1,...\mathbf{r}_N)$, so that the mean value of the value $Q(\mathbf{r};\mathbf{r}_1,...\mathbf{r}_N)$ at a point $\mathbf{r}$ is given by an integral:

$$\langle Q \rangle(\mathbf{r}) = \int Q(\mathbf{r};\mathbf{r}_1,...\mathbf{r}_N)P(\mathbf{r}_1,...\mathbf{r}_N)d\mathbf{r}_1 \cdot ... \cdot d\mathbf{r}_N.$$

In the following, we will assume the distribution of disks to be random and on average homogeneous. The disks cannot overlap, so, in general, $P(\mathbf{r}_1,...\mathbf{r}_N) = 0$ at $|\mathbf{r}_i - \mathbf{r}_j| < 2R$. We will not take this restriction into account, since its consideration leads to corrections quadratic in the small parameter $R^2 n_D$. This means that $P(\mathbf{r}_1,...\mathbf{r}_N) = p(\mathbf{r}_1) \cdot ... \cdot p(\mathbf{r}_N)$, where $p(\mathbf{r}_i) = 1/S$. Multiply the x-component of equation (1) by the function $H$ and average the result over the positions of all disks. This gives:

$$e\langle H E_x \rangle = m\nu \text{div}\langle (H\nabla V_x) \rangle - m\nu \langle \nabla V_x \cdot \nabla H \rangle. \tag{D2}$$

It is not difficult to show (see, e.g.,[39]) that $\text{div}\langle(H\nabla V_x)\rangle = \Delta\langle V_x\rangle$. *In the case of a homogeneous disk distribution* $\Delta\langle V_x\rangle = 0$. Taking advantage of the symmetry of the integrand with respect to the permutations of disks, we obtain from (D1) and (D2):

$$e\langle H\,\mathrm{E}_x\rangle = -m\nu n_D \int \delta(R - |\mathbf{r} - \mathbf{r}_1|)\nabla\langle V_x\rangle_1 \cdot \mathbf{e}_1 d\mathbf{r}_1, \quad \mathbf{e}_1 = \frac{\mathbf{r} - \mathbf{r}_1}{|\mathbf{r} - \mathbf{r}_1|}, \tag{D3}$$

where $\langle V_x\rangle_1$ is the average, assuming one disk is fixed:

$$\langle V_x\rangle_1 = \int V_x(\mathbf{r};\mathbf{r}_1,\ldots\mathbf{r}_N)\frac{d\mathbf{r}_2 \cdot \ldots \cdot d\mathbf{r}_N}{S^{N-1}}.$$

It is clear that $\langle V_x\rangle_1$ is a function of the difference $\mathbf{r} - \mathbf{r}_1$, which allows us to go from integration over $\mathbf{r}_1$ in (D3) to integration over $\boldsymbol{\rho} = \mathbf{r} - \mathbf{r}_1$. Taking into account the $\delta$-function, we obtain:

$$e\langle H\,\mathrm{E}_x\rangle = m\nu n_D R\int \frac{\partial\langle V_x\rangle_1}{\partial\rho}d\theta. \tag{D4a}$$

It comes out the same way:

$$e\langle H\,\mathrm{E}_y\rangle = m\nu n_D R\int \frac{\partial\langle V_y\rangle_1}{\partial\rho}d\theta. \tag{D4b}$$

Using now the relations (2), we find:

$$\begin{aligned} e\langle H\,\mathrm{E}_x\rangle &= m\nu R n_D \oint_\Gamma \langle\Omega\rangle_1 \sin\theta d\theta, \\ e\langle H\,\mathrm{E}_y\rangle &= -m\nu R n_D \oint_\Gamma \langle\Omega\rangle_1 \cos\theta d\theta \end{aligned}. \tag{D5}$$

These equalities do not take into account the field inside the disks. For this field we have:

$$\begin{aligned} e\langle(1-H)\mathrm{E}_x\rangle &= -en_D \int h(R-\rho)\langle E_x\rangle_1 d\boldsymbol{\rho}, \\ e\langle(1-H)\mathrm{E}_y\rangle &= -en_D \int h(R-\rho)\langle E_y\rangle_1 d\boldsymbol{\rho} \end{aligned},$$

which by virtue of (4) can be written in the form

$$\begin{aligned} e\langle(1-H)\mathrm{E}_x\rangle &= m\nu n_D \int h(R-\rho)\frac{\partial\langle\Omega\rangle_1}{\partial y}d\boldsymbol{\rho}, \\ e\langle(1-H)\mathrm{E}_y\rangle &= -m\nu n_D \int h(R-\rho)\frac{\partial\langle\Omega\rangle_1}{\partial x}d\boldsymbol{\rho} \end{aligned}.$$

or, after simple transformations, in the form

$$e\langle(1-H)\mathrm{E}_x\rangle = -m\nu R n_D \oint_\Gamma R \frac{\partial\langle\Omega\rangle_1}{\partial\rho}\sin\theta d\theta,$$

$$e\langle(1-H)\mathrm{E}_y\rangle = m\nu R n_D \oint_\Gamma R \frac{\partial\langle\Omega\rangle_1}{\partial\rho}\cos\theta d\theta$$

(D6)

Adding (D5) and (D6) and taking into account that $\langle \mathrm{E}_x\rangle = -U_x/L$ and $\langle \mathrm{E}_y\rangle = -U_y/L$, for $B \neq 0$, we obtain the equations (10).

From symmetry considerations it follows that the integrals in the right parts of (D4), and hence in the right parts of (15) and (16), are proportional to the mean flow velocity $(\overline{V}_x = \langle V_x\rangle, \overline{V}_y = \langle V_y\rangle)$, so that the right parts of Equations (10) take the form:

$$m\nu R n_D \oint_\Gamma \left(R\frac{\partial\langle\Omega\rangle_1}{\partial r} - \langle\Omega\rangle_1\right)\sin\theta d\theta = \frac{m\overline{V}_x}{\tau} + m\delta\omega_c \overline{V}_y,$$

$$-m\nu R n_D \oint_\Gamma \left(R\frac{\partial\langle\Omega\rangle_1}{\partial r} - \langle\Omega\rangle_1\right)\cos\theta d\theta = \frac{m\overline{V}_x}{\tau} - m\delta\omega_c \overline{V}_x$$

(D7)

The relaxation time $\tau$ and the "shift of the cyclotron frequency" $\delta\omega_c$ are to be determined.

Now we need to derive equations for the values $\langle \mathbf{V}\rangle_1$, $\langle \mathbf{E}\rangle_1$ and $\langle \Omega\rangle_1$. For this purpose, we fix one of the disks and, multiplying equation (1) by $H$, we average it over the positions of the other disks. Then, for the average local field in the flow region, we obtain (again, we temporarily assume $B = 0$):

$$e\langle H\mathrm{E}_x\rangle_1 = m\nu\Delta\langle V_x\rangle_1 - m\nu n_D \int \delta(R-|\mathbf{r}-\mathbf{r}_2|)\nabla\langle V_x\rangle_{1,2}\cdot \mathbf{e}_2 d\mathbf{r}_2,$$

$$e\langle H\mathrm{E}_y\rangle_1 = m\nu\Delta\langle V_y\rangle_1 - m\nu n_D \int \delta(R-|\mathbf{r}-\mathbf{r}_2|)\nabla\langle V_y\rangle_{1,2}\cdot \mathbf{e}_2 d\mathbf{r}_2,$$

where $\langle Q\rangle_{1,2}$ is the average provided that the positions of the two disks are fixed:

$$\langle Q\rangle_{1,2} = \int Q(\mathbf{r},\mathbf{r}_1,\mathbf{r}_2;\mathbf{r}_3,\ldots,\mathbf{r}_N)\frac{d\mathbf{r}_3\cdot\ldots\cdot d\mathbf{r}_N}{S^{N-2}}.$$

For the field in the area occupied by disks, we obtain similarly to (D6)

$$e\langle(1-H)\mathrm{E}_x\rangle_1 = n_D \int \Xi(R-|\mathbf{r}-\mathbf{r}_2|)\langle E_x\rangle_{1,2} d\mathbf{r}_2,$$

and for the mean total field we will have ($B \neq 0$):

$$e\langle E_x\rangle_1 = m\nu\Delta\langle V_x\rangle_1 - m\omega_c\langle V_y\rangle_1 - m\nu Rn_D\oint_\Gamma\left(R\frac{\partial\langle\Omega\rangle_{12}}{\partial r}-\langle\Omega\rangle_{12}\right)\sin\theta d\theta,$$

$$e\langle E_y\rangle_1 = m\nu\Delta\langle V_y\rangle_1 + m\omega_c\langle V_x\rangle_1 + m\nu Rn_D\oint_\Gamma\left(R\frac{\partial\langle\Omega\rangle_{12}}{\partial r}-\langle\Omega\rangle_{12}\right)\cos\theta d\theta$$

Finally, assuming, following [39], that the integrals in these equations are expressed over $\langle V_x\rangle_1$ and $\langle V_y\rangle_1$ in the same way as the analogous integrals in (D7) are expressed over $\overline{V}_x$ and $\overline{V}_y$, we obtain equation (11) of the main text (in it the angle brackets and inlex 1 for the velocity and electric field vectors are omitted). The point of the made assumptions is the idea to imagine the second disk, located in the R-neighborhood of the point $\mathbf{r}$, being in a stream flowing at a large distance from it with velocity $\langle V\rangle_1(\mathbf{r}-\mathbf{r}_1)$, rather than $\langle V\rangle$. In [39] the corresponding calculations are presented.

# Appendix E
## System of equations for distribution function in three-harmonic approximation

In this section, we present the detailed equations for the amplitudes of the angular harmonics of the truncated distribution function, which follow from the kinetic equation.

Substituting expression (49) into the kinetic equation, we obtain a system of equations:

$$\frac{v}{2}\frac{\partial f_{2c}}{\partial x}+\frac{v}{2}\frac{\partial f_{2s}}{\partial y}-eE_x\,v\,f_F'+\omega_c f_{1s}=0,\quad -\frac{v}{2}\frac{\partial f_{2c}}{\partial y}+\frac{v}{2}\frac{\partial f_{2s}}{\partial x}-eE_y\,v\,f_F'-\omega_c f_{1c}=0,\quad \frac{\partial f_{1c}}{\partial x}+\frac{\partial f_{1s}}{\partial y}=0, \quad (E1)$$

$$\frac{v}{2}\left(\frac{\partial f_{1c}}{\partial x}-\frac{\partial f_{1s}}{\partial y}+\frac{\partial f_{3c}}{\partial x}+\frac{\partial f_{3s}}{\partial y}\right)+2\omega_c f_{2s}+\frac{1}{\tau_2}f_{2c}=0,\quad \frac{v}{2}\left(\frac{\partial f_{1c}}{\partial y}+\frac{\partial f_{1s}}{\partial x}-\frac{\partial f_{3c}}{\partial y}+\frac{\partial f_{3s}}{\partial x}\right)-2\omega_c f_{2c}+\frac{1}{\tau_2}f_{2s}=0,$$

$$f_{3c}=\frac{l_3}{2(1+\beta_3^2)}\left(-\frac{\partial f_{2c}}{\partial x}+\frac{\partial f_{2s}}{\partial y}+\beta_3\frac{\partial f_{2c}}{\partial y}+\beta_3\frac{\partial f_{2s}}{\partial x}\right),\quad f_{3s}=-\frac{l_3}{2(1+\beta_3^2)}\left(\frac{\partial f_{2s}}{\partial x}+\frac{\partial f_{2c}}{\partial y}+\beta_3\frac{\partial f_{2c}}{\partial x}-\beta_3\frac{\partial f_{2s}}{\partial y}\right),$$

We can exclude the third-order harmonics $f_{3s}$ and $f_{3c}$ from these equations:

$$\frac{l_2}{2}\frac{\partial f_{1c}}{\partial x}-\frac{l_2}{2}\frac{\partial f_{1s}}{\partial y}-\frac{l_2 l_3}{4(1+\beta_3^2)}(\Delta f_{2c}-\beta_3\Delta f_{2s})+\beta f_{2s}+f_{2c}=0,\quad \beta_3=3\omega_c\tau_3,\quad l_n=v\tau_n, \quad (E2)$$

$$\frac{l_2}{2}\frac{\partial f_{1c}}{\partial y}+\frac{l_2}{2}\frac{\partial f_{1s}}{\partial x}-\frac{l_2 l_3}{4(1+\beta_3^2)}(\Delta f_{2s}+\beta_3\Delta f_{2c})-\beta f_{2c}+f_{2s}=0$$

Note that, according to our condition of instantaneous relaxation of all harmonics beginning with the fourth harmonic, in the expressions for $f_{2c}$ and $f_{2s}$ there is no contribution of the fourth harmonic.

According to the definitions (37) we have

$$\Pi_{xx} = -\Pi_{yy} = \frac{\pi m}{n_0} \int f_{2c} \, v^2 \frac{p\,dp}{(2\pi\hbar)^2}, \quad \Pi_{xy} = \Pi_{yx} = \frac{\pi m}{n_0} \int f_{2s} \, v^2 \frac{p\,dp}{(2\pi\hbar)^2}, \quad (E3)$$

$$V_x = \frac{2\pi}{n_0} \int f_{1c} \, v \frac{p\,dp}{(2\pi\hbar)^2}, \quad V_y = \frac{2\pi}{n_0} \int f_{1s} \, v \frac{p\,dp}{(2\pi\hbar)^2}$$

Multiplying equations (E2) by $V_i$ and $v_i v_k$ and integrating, we obtain the equations (50).

# Appendix F
## Boundary conditions at edges of smooth disks

Below we derive the boundary conditions at the disc edges for the momentum flux tensor calculated on the truncated three-harmonic distribution function.

Condition (57) on the distribution function can be rewritten as:

$$[1-(-1)^n \cos 2n\theta] f_{nc} - (-1)^n \sin 2n\theta \, f_{ns} = 0, \quad (F1)$$
$$-(-1)^n \sin 2n\theta \, f_{nc} + [1+(-1)^n \cos 2n\theta] f_{ns} = 0$$

These equations connect the functions $f_{nc}(\varepsilon, R, \theta)$ and $f_{ns}(\varepsilon, R, \theta)$:

$$f_{nc} \sin n\theta = f_{ns} \cos n\theta, \, n=2k; \quad f_{nc} \cos n\theta = -f_{ns} \sin n\theta, \, n=2k+1. \quad (F2)$$

For $n=2$ we have from here $f_{2c} \sin 2\theta = f_{2s} \cos 2\theta$, which is equivalent to the hydrodynamic condition $\Pi_{r\theta}\big|_{r=R} = 0$. For the third harmonic, relations (D2) give $f_{3c} \cos 3\theta + f_{3s} \sin 3\theta = 0$, whence, taking into account the first of equations (50), we obtain:

$$\left( -\frac{\partial f_{2c}}{\partial x} + \frac{\partial f_{2s}}{\partial y} + \beta_3 \frac{\partial f_{2c}}{\partial y} + \beta_3 \frac{\partial f_{2s}}{\partial x} \right)_{r=R} \cos 3\theta - \left( \frac{\partial f_{2s}}{\partial x} + \frac{\partial f_{2c}}{\partial y} + \beta_3 \frac{\partial f_{2c}}{\partial x} - \beta_3 \frac{\partial f_{2s}}{\partial y} \right)_{r=R} \sin 3\theta = 0,$$

or, equivalently:

$$\left( -\frac{\partial \Pi_{xx}}{\partial x} + \frac{\partial \Pi_{xy}}{\partial y} + \beta_3 \frac{\partial \Pi_{xx}}{\partial y} + \beta_3 \frac{\partial \Pi_{xy}}{\partial x} \right)_{r=R} \cos 3\theta - \left( \frac{\partial \Pi_{xy}}{\partial x} + \frac{\partial \Pi_{xx}}{\partial y} + \beta_3 \frac{\partial \Pi_{xx}}{\partial x} - \beta_3 \frac{\partial \Pi_{xy}}{\partial y} \right)_{r=R} \sin 3\theta = 0.$$

From these equations rewritten in the polar coordinates, using the formulas:

$$\Pi_{rr} = \Pi_{xx} \cos 2\theta + \Pi_{xy} \sin 2\theta, \quad \Pi_{r\theta} = -\Pi_{xx} \sin 2\theta + \Pi_{xy} \cos 2\theta,$$
$$\Pi_{xy} = \Pi_{rr} \sin 2\theta + \Pi_{r\theta} \cos 2\theta, \quad \Pi_{xx} = \Pi_{rr} \cos 2\theta - \Pi_{r\theta} \sin 2\theta, \quad (F3)$$

we obtain the second of condition (58).

We emphasize that the first condition (58) is exact, while the second condition is approximate, since in its derivation in expressions (C1) for $f_{3c}$ and $f_{3s}$ the contribution of the fourth harmonic has been omitted.

# Appendix G
## Calculation of tensor $\Pi_{\alpha\beta}$ in three-harmonic approximation

In this section, we construct the solution of the hydrodynamic like equations for flow function $\psi(\mathbf{r})=2\text{Re}[\chi(r)e^{i\theta}]$, electrostatic potential $\Phi(\mathbf{r})=2\text{Re}[\phi(r)e^{i\theta}]$, and the momentum flux $\Pi_{ik}(\mathbf{r})=2\text{Re}[Q_{ik}(r)e^{i\theta}]$ within the three-harmonic approximation. The differential operator on the left-hand side of the first equation (67b) is zeroed by the two functions $1/r$ and $1/r^3$. Thus, we seek a solution in the form:

$$Q_{r\theta} = \frac{a}{r} + \frac{g(r)}{r^3},$$

where $a$ is a constant and $g(r)$ is a new unknown function. The equation for $g$ is:

$$\frac{\partial^2 g}{\partial r^2} - \frac{1}{r}\frac{\partial g}{\partial r} = r\frac{\partial F}{\partial r}.$$

It contains only derivatives of $g(r)$ and is easily solved. As a result we have:

$$Q_{r\theta} = \frac{b}{r} + \frac{c}{r^3} + \frac{1}{r^3}\int rF(r)dr = \frac{b}{r} + \frac{c}{r^3} + \frac{1}{r^3}\int\left[2ir^2\phi - \frac{m}{\tau}r^2\frac{\partial(r\chi)}{\partial r}\right]dr$$

$$= \frac{b}{r} + \frac{c}{r^3} - \frac{m}{\tau}\chi + \frac{2}{r^3}\int r^2(i\phi + m\chi/\tau)dr$$

Substituting here expressions (62) and (63) and taking into account that $\int x^2 K_1(x)dx = -x^2 K_2(x)$, we obtain for the tensor $Q_{\alpha\beta}$:

$$Q_{r\theta} = \frac{b}{r} + \frac{c}{r^3} + \frac{ir}{2}\left(\tilde{\alpha} + \frac{im}{\tau}\alpha\right) + \frac{i\tilde{\delta}}{r} - \frac{m}{\tau}[\gamma_+ K_1(\sqrt{\kappa_+}r) + \gamma_- K_1(\sqrt{\kappa_-}r)]$$

$$-\frac{2}{r}\left[\frac{1}{\sqrt{\kappa_+}}\left(i\tilde{\gamma}_+ + \frac{m}{\tau}\gamma_+\right)K_2(\sqrt{\kappa_+}r) + \frac{1}{\sqrt{\kappa_-}}\left(i\tilde{\gamma}_- + \frac{m}{\tau}\gamma_-\right)K_2(\sqrt{\kappa_-}r)\right],$$

$$Q_{rr} = -\frac{ib}{r} + \frac{ic}{r^3} + \phi + \frac{2i}{r^3}\int r^2(i\phi + m\chi/\tau)dr$$

$$= -\frac{ib}{r} + \frac{ic}{r^3} + \frac{ir}{2}\left(\tilde{\alpha} + \frac{im}{\tau}\alpha\right) + \frac{im\delta}{\tau r} + \tilde{\gamma}_+ K_1(\sqrt{\kappa_+}r) + \tilde{\gamma}_- K_1(\sqrt{\kappa_-}r)$$

$$-\frac{2i}{r}\left[\frac{1}{\sqrt{\kappa_+}}\left(i\tilde{\gamma}_+ + \frac{m}{\tau}\gamma_+\right)K_2(\sqrt{\kappa_+}r) + \frac{1}{\sqrt{\kappa_-}}\left(i\tilde{\gamma}_- + \frac{m}{\tau}\gamma_-\right)K_2(\sqrt{\kappa_-}r)\right]. \tag{G1}$$

The last two formulas in Eqs. (50) In radial variables r and $\theta$ take the form:

$$\Pi_{r\theta} = -m\nu\left(\frac{1}{r}\frac{\partial V_r}{\partial \theta} + \frac{\partial V_\theta}{\partial r} - \frac{V_\theta}{r}\right) - m\nu\beta_2\left(\frac{\partial V_r}{\partial r} - \frac{1}{r}\frac{\partial V_\theta}{\partial \theta} - \frac{V_r}{r}\right)$$
$$+ \frac{\nu\tau_3}{1+\beta_3^2}(1-\beta_2\beta_3)\left(\Delta\Pi_{r\theta} - \frac{4\Pi_{r\theta}}{r^2} + \frac{4\partial\Pi_{rr}}{r^2\partial\theta}\right) + \frac{\nu\tau_3}{1+\beta_3^2}(\beta_2+\beta_3)\left(\Delta\Pi_{rr} - \frac{4\Pi_{rr}}{r^2} - \frac{4\partial\Pi_{r\theta}}{r^2\partial\theta}\right),$$
$$\Pi_{rr} = -m\nu\left(\frac{\partial V_r}{\partial r} - \frac{1}{r}\frac{\partial V_\theta}{\partial \theta} - \frac{V_r}{r}\right) + m\nu\beta_2\left(\frac{1}{r}\frac{\partial V_r}{\partial \theta} + \frac{\partial V_\theta}{\partial r} - \frac{V_\theta}{r}\right)$$
$$+ \frac{\nu\tau_3}{1+\beta_3^2}(1-\beta_2\beta_3)\left(\Delta\Pi_{rr} - \frac{4\Pi_{rr}}{r^2} - \frac{4\partial\Pi_{r\theta}}{r^2\partial\theta}\right) - \frac{\nu\tau_3}{1+\beta_3^2}(\beta_2+\beta_3)\left(\Delta\Pi_{r\theta} - \frac{4\Pi_{r\theta}}{r^2} + \frac{4\partial\Pi_{rr}}{r^2\partial\theta}\right)$$
(G2)

From (G1) and (G2), using the definitions $V_r = --\partial\psi/r\partial\theta$, $V_\theta = \partial\psi/\partial r$ and Eqs. (62), (63), it can be seen that the tensor should not contain linear in r terms, whence it follows $\tilde{\alpha} = -im\alpha/\tau$. Moreover, we can also see that there are no terms proportional to $r^{-1}$, which gives $b = m\delta/\tau$ one more connection between the constants of $\tilde{\delta} = im\delta/\tau$. Finally, the terms proportional to $r^{-3}$, enter only in the hydrodynamic part of the tensor and have the form $-4m\nu\delta(1+i\beta)/r^3$ in the case $\Pi_{r\theta}$ and $-4m\nu\delta(i-\beta)/r^3$ in the case of $\Pi_{rr}$. Comparing these relations with Eq. (G1), we obtain $c = -4m\nu(1+i\beta_2)\delta$.

Thus, all constants of integration are expressed through four quantities $\alpha, \delta, \gamma_+$ and $\gamma_-$, which can be found from the four boundary conditions: the three ones on the edges of the disk and the one at infinity, $\rho \to \infty$. In the future it will be convenient to use the dimensionless radius $\rho = r/R$, denote $\alpha R$ by $\alpha$, $\delta/R$ by $\delta$ and introduce the parameters $\varepsilon_- = \sqrt{\kappa_-}R \ll 1$ and $\varepsilon_+ = \sqrt{\kappa_+}R$. The tensor $Q_{\alpha\beta}$ is written as follows:

$$Q_{r\theta} = -\frac{4m\nu\delta}{R^2\rho^3}(1+i\beta_2) - \frac{m}{\tau}[\gamma_+ K_1(\varepsilon_+\rho) + \gamma_- K_1(\varepsilon_-\rho)]$$
$$- \frac{2}{\rho}\left[\frac{1}{\varepsilon_+}\left(i\tilde{\gamma}_+ + \frac{m}{\tau}\gamma_+\right)K_2(\varepsilon_+\rho) + \frac{1}{\varepsilon_-}\left(i\tilde{\gamma}_- + \frac{m}{\tau}\gamma_-\right)K_2(\varepsilon_-\rho)\right],$$
$$Q_{rr} = -i\frac{4m\nu\delta}{R^2\rho^3}(1+i\beta_2) + \tilde{\gamma}_+ K_1(\varepsilon_+\rho) + \tilde{\gamma}_- K_1(\varepsilon_-\rho)$$
$$- \frac{2i}{\rho}\left[\frac{1}{\varepsilon_+}\left(i\tilde{\gamma}_+ + \frac{m}{\tau}\gamma_+\right)K_2(\varepsilon_+\rho) + \frac{1}{\varepsilon_-}\left(i\tilde{\gamma}_- + \frac{m}{\tau}\gamma_-\right)K_2(\varepsilon_-\rho)\right]$$
(G3)

The boundary condition $V_r(1) = 0$ is equivalent, as it was in hydrodynamics, to the condition $\chi(1) = 0$, so from (62) we have

$$\delta = -\alpha - \gamma_+ K_1(\varepsilon_+) - \gamma_- K_1(\varepsilon_-).$$
(G4)

The conditions $V_x(\infty) = V_x$, $V_y(\infty) = V_y$ yield:

$$\alpha = \frac{RV_y}{2} + i\frac{RV_x}{2} \equiv \alpha_1 + i\alpha_2.$$
(G5)

Since at $B \to 0$ and the values $\tilde{\gamma}_-$ and $\gamma_+$ turn to zero, it is reasonable to use the coefficients $\gamma_-$ and $\tilde{\gamma}_+$. The condition $Q_{r\theta}(1) = 0$ gives:

$$-\frac{4mv\delta}{R^2}(1+i\beta_2) - \frac{m}{\tau}[\gamma_+ K_1(\varepsilon_+) + \gamma_- K_1(\varepsilon_-)]$$

$$-2\left[\frac{1}{\varepsilon_+}\left(i\tilde{\gamma}_+ + \frac{m}{\tau}\gamma_+\right)K_2(\varepsilon_+) + \frac{1}{\varepsilon_-}\left(i\tilde{\gamma}_- + \frac{m}{\tau}\gamma_-\right)K_2(\varepsilon_-)\right] = 0$$ (G6)

Finally, it is necessary to satisfy the second boundary condition (58), which for the first harmonic is written as:

$$\left[-\frac{\partial Q_{rr}}{\partial \rho} + 2Q_{rr} + \beta_3\left(\frac{\partial Q_{r\theta}}{\partial \rho} + iQ_{rr}\right)\right]_{\rho=1} = 0,$$

or, taking into account equations (67) and the condition $\chi(1) = 0$, in the form:

$$\left[2(2+i\beta_3)Q_{rr} - \frac{\partial \Phi}{\partial \rho} + i\beta_3\left(\Phi + i\frac{m}{\tau}\frac{\partial \chi}{\partial \rho}\right)\right]_{\rho=1} = 0.$$ (G7)

Thus, we have the following system of equations for finding the values $\gamma_-$ and $\tilde{\gamma}_+$:

$$-\frac{4mv\delta}{R^2}(1+i\beta_2) - \frac{m}{\tau}[\gamma_+ K_1(\varepsilon_+) + \gamma_- K_1(\varepsilon_-)] - 2i\left[\frac{1}{\varepsilon_+}\left(\tilde{\gamma}_+ - i\frac{m}{\tau}\gamma_+\right)K_2(\varepsilon_+) + \frac{1}{\varepsilon_-}\left(\tilde{\gamma}_- - i\frac{m}{\tau}\gamma_-\right)K_2(\varepsilon_-)\right] = 0$$

$$\left[2(2+i\beta_3)Q_{rr} - \frac{\partial \Phi}{\partial \rho} + i\beta_3\left(\Phi + i\frac{m}{\tau}\frac{\partial \chi}{\partial \rho}\right)\right]_{\rho=1} = 0,$$ (G8)

where:

$$Q_{rr}(1) = -i\frac{4mv\delta}{R^2}(1+i\beta_2) + \tilde{\gamma}_+ K_1(\varepsilon_+) + \tilde{\gamma}_- K_1(\varepsilon_-) + 2\left[\frac{1}{\varepsilon_+}\left(\tilde{\gamma}_+ - i\frac{m}{\tau}\gamma_+\right)K_2(\varepsilon_+) + \frac{1}{\varepsilon_-}\left(\tilde{\gamma}_- - i\frac{m}{\tau}\gamma_-\right)K_2(\varepsilon_-)\right].$$ (G9)

$$\delta = -\alpha - \gamma_+ K_1(\varepsilon_+) - \gamma_- K_1(\varepsilon_-), \quad \frac{\partial \chi}{\partial \rho} = 2\alpha - \gamma_+ \varepsilon_+ K_0(\varepsilon_+) - \gamma_- \varepsilon_- K_0(\varepsilon_-)$$

$$\Phi(1) = -2i\frac{m}{\tau}\alpha + \left(\tilde{\gamma}_+ - i\frac{m}{\tau}\gamma_+\right)K_1(\varepsilon_+) + \left(\tilde{\gamma}_- - i\frac{m}{\tau}\gamma_-\right)K_1(\varepsilon_-),$$

$$\frac{\partial \Phi(1)}{\partial \rho} = -2i\frac{m}{\tau}\alpha - \left(\tilde{\gamma}_+ - i\frac{m}{\tau}\gamma_+\right)K_1(\varepsilon_+) - \left(\tilde{\gamma}_- - i\frac{m}{\tau}\gamma_-\right)K_1(\varepsilon_-) - \tilde{\gamma}_+ \varepsilon_+ K_0(\varepsilon_+) - \tilde{\gamma}_- \varepsilon_- K_0(\varepsilon_-),$$

$$\Phi(1) + i\frac{m}{\tau}\frac{\partial \chi(1)}{\partial \rho} = \left(\tilde{\gamma}_+ - i\frac{m}{\tau}\gamma_+\right)K_1(\varepsilon_+) + \left(\tilde{\gamma}_- - i\frac{m}{\tau}\gamma_-\right)K_1(\varepsilon_-) - i\frac{m}{\tau}\gamma_+ \varepsilon_+ K_0(\varepsilon_+) - i\frac{m}{\tau}\gamma_- \varepsilon_- K_0(\varepsilon_-)$$

The formulas (G9) were derived using the identities $xK_1'(x) + K_1(x) = -xK_0(x)$ and $xK_2(x) = 2K_1(x) + xK_0(x)$. In this way, we have derived a close system of equations for the constants $\alpha$, $\delta$, $\gamma_\pm$, and $\tilde{\gamma}_\pm$.